\documentclass[11pt]{article}

\usepackage[usenames,dvipsnames,table]{xcolor}

\usepackage{jheppub} 

\usepackage{graphicx}
\usepackage{amsmath, amssymb}
\usepackage{placeins}
\usepackage[multiple]{footmisc}

\usepackage{enumerate}

\newcommand{\be}{\begin{eqnarray}}
\newcommand{\ee}{\end{eqnarray}}
\newcommand{\bea}{\begin{eqnarray}}
\newcommand{\eea}{\end{eqnarray}}

\newcommand{\nn}{\nonumber}
\newcommand{\bn}{\begin{enumerate}}
\newcommand{\en}{\end{enumerate}}

\def\Tr{\mathop{\mathrm{Tr}}\nolimits}

\def\e{\epsilon}

\def\e{\textrm{e}}

\newcommand{\udl}[1]{\mathrm{d} #1 \,}

\newcommand{\qfac}[1]{\left( #1; q \right)_\infty}

\newcommand{\thetafunc}[1]{\gt\left( #1;q\right)}

\def\ga{\alpha}
\def\gb{\beta}
\def\gc{\gamma}

\def\gd{\delta}

\def\gt{\theta}

\def\gs{\sigma}

\def\gr{\rho}

\def\Gp{\Phi}

\title{New 2d $\mathcal{N}=(0,2)$ dualities from four dimensions}

\preprint{}

\author[a]{Matteo Sacchi}

\affiliation[a]{Dipartimento di Fisica, Universit\`a di Milano-Bicocca \& INFN, Sezione di Milano-Bicocca,
I-20126 Milano, Italy}

\emailAdd{m.sacchi13@campus.unimib.it}

\abstract{
We propose some new infra-red dualities for $2d$ $\mathcal{N}=(0,2)$ theories. 
The first one relates a $USp(2N)$ gauge theory with one antisymmetric chiral, four fundamental chirals and $N$ Fermi singlets to a Landau--Ginzburg model of $N$ Fermi and $6N$ chiral fields with cubic interactions. The second one relates $SU(2)$ linear quiver gauge theories of arbitrary length $N-1$ with the addition of $N$ Fermi singlets for any non-negative integer $N$.
They can be understood as a generalization of the duality between an $SU(2)$ gauge theory with four fundamental chirals and a Landau--Ginzburg model of one Fermi and six chirals with a cubic interaction. 
We derive these dualities from already known $4d$ $\mathcal{N}=1$ dualities by compactifications on $\mathbb{S}^2$ with suitable topological twists and we further test them by matching anomalies and elliptic genera.
We also show how to derive them by iterative applications of some more fundamental dualities, in analogy with similar derivations for parent dualities in three and four dimensions.
}

\begin{document} 

\maketitle
\flushbottom

\section{Introduction}

Among the most fascinating phenomena that may characterize the low energy dynamics of a quantum field theory are infra-red dualities. This occurs when two different microscopic theories become equivalent in the infra-red (IR). A vast amount of such dualities have been discovered over the years since the first example of Seiberg duality in four dimensions \cite{Seiberg:1994pq}\footnote{See for example \cite{Hori:2006dk,Hori:2011pd} for two-dimensional versions of Seiberg duality with $\mathcal{N}=(2,2)$ supersymmetry.}, especially for theories with supersymmetry. These theories possess indeed several protected quantities that can be used to test dualities, such as partition functions on different backgrounds which have been compute exactly using localization techniques (for a review see \cite{Pestun:2016zxk}, especially contribution \cite{Benini:2016qnm} for localization in two dimensions). 

Nevertheless, a complete understanding of an organizing principle behind the currently known IR dualities in two, three and four dimensions hasn't been achieved yet. One possible approach towards this goal resides in the study of dimensional reductions of dualities. One can indeed hope that all of the dualities in lower dimensions can be derived from a small, restricted set of dualities in higher dimensions. Remarkable results have been obtained over the last few years in this direction. 

A different point of view on the understanding of the dimensional reduction limit of IR dualities is that this can be used to derive new dualities from already known ones. More precisely, we can start from a known duality in $d$ dimensions, compactify both of the dual theories on a $(d-d')$-dimensional manifold and flow to energies much smaller than the compactification scale. In this way we obtain two theories in $d'$ dimensions and we can ask ourselves if they are still dual or not. In some cases, when enough insight is gained about the dimensional reduction limit, one can even push this approach further and try to reverse the logic. Namely, we can start from a duality in lower dimensions and use it to guess a still unknown parent duality in higher dimensions. This should not be intended as an attempt of reversing the RG flow, but just as a hint for the existence of the higher dimensional duality.

There are several subtleties that one has to take into account when studying the dimensional reduction of a duality and in general it may just happen that the two lower dimensional theories are not dual. This is due to the fact that two different limits are involved in the dimensional reduction and issues of order of limits are typically involved. The first limit consists of flowing to low energies while keeping the compactification radius $r$ fixed. Here is where the duality holds and we expect the two $d$-dimensional theories to flow to the same fixed point. The second limit is the strict dimensional reduction limit $r\to 0$ while keeping the energy scale fixed. Taking the two limits in this order would give us the lower dimensional version of the fixed point theory of the original $d$-dimensional dual theories. If we instead take the limit $r\to 0$ first, we obtain two lower dimensional theories that we can conjecture being two different UV description of the aforementioned $d'$-dimensional fixed point theory. Thus, we understand that this conjecture is true and that the duality survives the dimensional reduction only if the two limits commute, but this is not always true.

This problem has so far been understood at a different level depending on the setup considered. The most understood case is the dimensional reduction of $4d$ $\mathcal{N}=1$ dualities on $\mathbb{S}^1$, giving dualities between $3d$ $\mathcal{N}=2$ theories \cite{Aharony:2013dha,Aharony:2013kma}. A crucial role is played here by monopole operators, in the sense that the two UV theories in three dimensions are dual to each other provided that they are supplemented by some monopole superpotential. These additional superpotential terms also explain a mismatch of symmetries that we can have between the $4d$ and the $3d$ theories. Indeed, four-dimensional theories typically possess anomalous $U(1)$ axial symmetries, which are not anomalous in three dimensions. The monopole superpotential has precisely the effect of breaking the symmetries of the three-dimensional theory that were anomalous in $4d$.

Another set-up that was analyzed in details is the dimensional reduction of $3d$ $\mathcal{N}=2$ theories on $\mathbb{S}^1$, giving dualities between $2d$ $\mathcal{N}=(2,2)$ theories \cite{Aganagic:2001uw,Aharony:2017adm}. In this case there are subtleties related to the fact that the resulting two-dimensional theories may have a non-compact target space \cite{Aharony:2016jki}. Indeed, in $2d$ the ground state can explore the entire moduli space of the theory because of quantum fluctuations, so that we can't just focus on a single region of it. Moreover, the metric on the target space, which is not protected by supersymmetry, is classically marginal in two dimensions. Consequently, in order to claim for a duality we need a complete knowledge of the target space of the theories at the quantum level, which is in general extremely difficult to achieve. This problem doesn't appear when the theories have a compact target space or when massive deformations are turned on, since these have the effect of lifting the vacua of the theory, leaving just discrete isolated vacua, but in the non-compact case it is not guaranteed that the duality survives when massive deformations are switched off. 

In this paper we are interested in the dimensional reduction of $4d$ $\mathcal{N}=1$ dualities on $\mathbb{S}^2$ with a topological twist to give dualities between $2d$ $\mathcal{N}=(0,2)$ theories \cite{Gadde:2015wta}. There has been renewed interest in $2d$ $\mathcal{N}=(0,2)$ dualities after the discovery of the trialities of \cite{Gadde:2013lxa,Gadde:2014ppa} (see also \cite{Putrov:2015jpa,Gukov:2019lzi} for examples with a different amount of chiral supersymmetry) and in \cite{Gadde:2015wta} it was shown how to derive them by dimensional reduction of Seiberg duality in four dimensions. This type of dimensional reduction is characterized by many more subtleties than the aforementioned cases, some of them being the followings:
\begin{itemize}
\item Truncation to the zero modes. A single $4d$ theory compactified on $\mathbb{S}^2$ may lead in general to an infinite direct sum of $2d$ theories describing the various KK modes of the $4d$ fields. A prescription was given in \cite{Gadde:2015wta} for how to obtain a single $2d$ theory describing the zero modes only, which we are going to review.
\item Non-compact target space. Also in this set-up there could be problems related to the non-compactness of the target spaces of the resulting $2d$ theories. For this reason, all the dualities discussed in this paper should be more appropriately considered as dualities between mass deformed theories.
\item Anomalous vs.~non-anomalous global symmetries. Because of the different nature of anomalies in two and four dimensions, it may happen that a $U(1)$ global symmetry which was anomalous in $4d$ is not anomalous in $2d$. Typically the two-dimensional dualities obtained from four dimensions only hold provided that this symmetry is broken also in the $2d$ theory. This situation is reminiscent of the monopole superpotential that is non-perturbatively generated when going from $4d$ to $3d$, but it is still not clear what should cause such a symmetry breaking in $2d$. The majority of the examples we will discuss are not affected by such a problem.
\end{itemize}

On top of these issues, it may still happen that the resulting two-dimensional theories are not dual to each other and that standard tests of the duality, such as matching global symmetries, anomalies or elliptic genera, don't work. In this paper we are going to present some cases where the dimensional reduction seems to work, leading to new dualities for $2d$ $\mathcal{N}=(0,2)$ theories starting from known $4d$ $\mathcal{N}=1$ dualities. We will perform several tests of the proposed dualities and show how to derive them by iterative applications of some already known dualities.

The paper is organized as follows. In Section \ref{sec2} we review some known $2d$ $\mathcal{N}=(0,2)$ dualities that will play a role in our discussion and the prescription of \cite{Gadde:2015wta} for how to derive them from four dimensions. In Section \ref{sec3} we propose a new $2d$ duality for a $USp(2N)$ gauge theory with antisymmetric matter, which is obtained from a parent four-dimensional duality discussed in \cite{Csaki:1996eu}. In Section \ref{sec4} we propose an infinite dimensional family of dual frames made of $SU(2)$ linear quiver gauge theories of arbitrary length $N-1$ and with $N$ Fermi singlets, where $N$ is a non-negative integer number, which is derived from one of the $4d$ mirror-like dualities of \cite{Hwang:2020wpd}. In Appendix \ref{appA} we comment on the possibility of another duality for a $USp(2N)$ gauge theory with antisymmetric matter, but with a different number of fundamental matter fields than the one of Section \ref{sec3}. Finally, in Appendix \ref{appB} we summarize our conventions for the elliptic genus.

\section{Review material}
\label{sec2}

In this section we review some known results that will be important in our next discussion. We first describe some aspects of the $2d$ $\mathcal{N}=(0,2)$ duality between the $SU(2)$ gauge theory with 4 fundamental chirals and the Landau--Ginzburg (LG) model of one Fermi and 6 chirals with a cubic interaction proposed in \cite{Gadde:2015wta} and further analyzed in \cite{Dedushenko:2017osi}. Then, we briefly explain the prescription of \cite{Gadde:2015wta} for the reduction of $4d$ $\mathcal{N}=1$ theories to $2d$ $\mathcal{N}=(0,2)$ on $\mathbb{S}^2$ with a topological twist. We conclude the section revisiting the dimensional reduction of Intriligator--Pouliot duality \cite{Intriligator:1995ne} studied in \cite{Gadde:2015wta}, focusing in particular on the confining case.

\subsection{Duality for the $SU(2)$ gauge theory with 4 chirals}
\label{review1}

We consider the following $2d$ $\mathcal{N}=(0,2)$ duality \cite{Gadde:2015wta,Dedushenko:2017osi}:

\medskip
\noindent \textbf{Theory \boldmath$\mathcal{T}_{\text{A}}$}: $SU(2)$ gauge theory with four fundamental chiral fields $Q_a$ and no interaction\footnote{Recall that in a $2d$ $\mathcal{N}=(0,2)$ theory we can have $E$ and $J$ interactions between Fermi multiplets $\Psi^i$ and chiral multiplets $\Gp^a$ (see \cite{Tachikawa:2018sae} for a review). In all the examples we will consider in this paper, only $J$-interactions will be involved, which take the form
\be
\int\udl{\bar{\theta}^+}W\left(\Psi^i,\Gp^a\right)=\int\udl{\bar{\theta}^+}\Psi^i\,J_i\left(\Gp^a\right)\,,
\ee
where $J_i$ are generic holomorphic functions of the chiral multiplets $\Gp^a$ only. Hence, $W$ has to be a Fermi operator of $2d$ $\mathcal{N}=(0,2)$ R-charge one. By abuse of notation, we will call $W$ the ``superpotential" of the $2d$ theory.}
\be
W_{\mathcal{T}_{\text{A}}}=0\,.
\ee

\medskip
\noindent \textbf{Theory \boldmath$\mathcal{T}_{\text{B}}$}: LG model of one Fermi field $\Psi$ and six chiral fields $\Gp_{ab}$ for $a<b=1,\cdots,4$ with cubic interaction
\be
W_{\mathcal{T}_{\text{B}}}=\Psi\,\mathrm{Pf}\,\Gp\,.
\label{superpotTB}
\ee

The global symmetry of both of the theories is $SU(4)_u\times U(1)_s$\footnote{Throughout the paper we will label the factors in the global symmetry groups with the names we will use for the corresponding fugacities in the elliptic genus.}, under which the fields transform according to
\begin{table}[h]
\centering
\scalebox{1}{
\setlength{\extrarowheight}{1pt}
\begin{tabular}{c|cc|c}
{} & $SU(4)_u$ & $U(1)_s$ & $U(1)_{R_0}$ \\ \hline
$Q$ & $\bf 4$ & 1 & $0$ \\ \hline
$\Psi$ & $\bullet$ & $-4$ & $1$ \\
$\Gp$ & $\bf 6$ & 2 & $0$
\end{tabular}}
\end{table}

\noindent where we also introduced a possible choice of UV trial right-moving R-symmetry $U(1)_{R_0}$. When we flow to low energies this can mix with all the other abelian global symmetries of the theory and the exact superconformal one of the IR theory will take the form
\be
R=R_0+q_s\,R_s\,,
\label{trialRsymmetry}
\ee
where $q_s$ is the charge under $U(1)_s$ and $R_s$ is the mixing coefficient, which can be fixed with $c$-extremization \cite{Benini:2012cz}.

On top of matching global symmetries, there are several tests that we can perform for this duality. One consists of matching anomalies for the global symmetries. For both of the theories we find\footnote{We use conventions where the chirality matrix $\gc^3$ takes value $+1$ on right-handed fermions and $-1$ on left-handed fermions. For example
\be
\Tr\,\gc^3=n_{\text{chir}}-n_{\text{ferm}}-d_G\,,
\ee
where $n_{\text{chir}}$ is the number of chiral multiplets in the theory, $n_{\text{ferm}}$ is the number of Fermi multiplets and the last term is the contribution of the vector multiplet, with $d_G$ being the dimension of the adjoint representation of the gauge group $G$.}\footnote{We recall the Dynkin indices for fundamental, adjoint and antisymmetric representations of $SU(N)$
\be
T_{SU(N)}({\bf N})=\frac{1}{2}, \qquad T_{SU(N)}({\bf N^2-1})=N, \qquad T_{SU(N)}({\bf N(N-1)/2})=\frac{N-2}{2}\,,
\ee
and of $USp(2N)$
\be
T_{USp(2N)}({\bf N})=\frac{1}{2}, \qquad T_{USp(2N)}({\bf N(2N+1)})=N+1, \qquad T_{USp(2N)}({\bf N(2N-1)-1})=N-1\,.
\ee}
\be
\Tr\,\gc^3\,U(1)_s^2=8, \qquad \Tr\,\gc^3\,SU(4)_u^2=1\,.
\label{anomaliesfundduality}
\ee

Using the generic parametrization of the R-symmetry \eqref{trialRsymmetry}, we can compute the trial central charges of the dual theories and verify that they match
\be
c_R=3\Tr\,\gc^3\,U(1)_R^2=15-48R_s+24R_s^2,\qquad c_R-c_L=\Tr\,\gc^3=5\,.
\label{centrachargesfundduality}
\ee
Here we encounter a curious feature of this theory. If we try to extremize $c_R$ to find the value of $R_s$ corresponding to the superconformal R-charge we get $R_s=1$. Plugging this back into the trial central charges we obtain $c_R=-9$ and $c_L=-14$, which violate the unitarity bound. This signals that the way we implemented $c$-extremization was incorrect. There are two possible explanations for such a phenomenon: either the theory is SUSY breaking in the IR or it has a non-compact target space. For this particular duality, the failure of the naive application of $c$-extremization was interpreted in \cite{Dedushenko:2017osi} as due to the non-compactness of the target space. We can understand that the dual theories are not SUSY breaking from the fact that their elliptic  genera (see the next paragraph for their definitions) are non-vanishing\footnote{Since the elliptic genus is a refined version of the Witten index, we can figure out if the theory is SUSY breaking or not by computing it and verifying that it is non-zero. This may happen when the theory has too few fundamental chirals, since we don't have enough poles to make the integral \eqref{EGgeneral} in terms of which we can express the elliptic genus non-vanishing \cite{Gadde:2014ppa,Gadde:2013lxa}.}. Moreover, the elliptic genera are also divergent as we turn off flavor fugacities, which signals the presence of a non-compact direction in the target space. This non-compactness of the moduli space is already visible from a classical analysis. Indeed, from the equation of motion of the Fermi field $\Psi$ we can see that the classical moduli space of theory $\mathcal{T}_{\text{B}}$ is a complex hypersurface in $\mathbb{C}^6$ defined by the equation
\be
\mathrm{Pf}\,\Gp=0\,.
\label{eom}
\ee
When the target space is non-compact, one of the key assumptions of \cite{Benini:2012cz,Benini:2013cda} is violated.
In particular, what goes wrong in such a case is that there could be a non-holomorphic current for the flavor symmetry associated to such a non-compact direction, which can't mix with the R-symmetry current. When we write the trial central charge as in \eqref{centrachargesfundduality} we are instead allowing for such a mixing and this is what we are doing wrong in our naive application of $c$-extremization. In order to solve this problem and correctly apply $c$-extremization, we have to enforce in \eqref{centrachargesfundduality} that we have no mixing with any non-holomorphic flavor current arising from non-compact directions in the target space. At a pratical level, we have to identify which are the gauge invariant operators parametrizing such a direction in the moduli space and fix their R-charges to zero. In our particular case, the non-compact directions are parametrized by five out of the six chirals $\Gp_{ab}$ because of the constraint \eqref{eom}. Since these six chirals belong to the same representation of the $SU(4)_u$ global symmetry, we must require that they all have zero R-charge and this fixes the mixing coefficient to  $R_s=0$. Consequently, the R-charge of the fundamental chirals $Q_a$ on the side of theory $\mathcal{T}_{\text{A}}$ is fixed to zero, as the mesons $Q_aQ_b$ are mapped to the chirals $\Gp_{ab}$ across the duality. Hence, we get the central charges\footnote{The author is very grateful to the anonymous referee of JHEP for explaining this procedure for correctly extracting the conformal central charges of a theory with a non-compact target space.}\footnote{This result suggestes that the theories are flowing to an SCFT fixed point despite of the non-compact nature of the target space.}
\be
c_R=15,\qquad c_L=10\,.
\ee

Finally, we can match the elliptic genera of the two theories \cite{Gadde:2013wq,Gadde:2013ftv,Benini:2013nda,Benini:2013xpa} (see also Appendix \ref{appB} for our conventions)
\be
\mathcal{I}_{\mathcal{T}_{\text{A}}}=\oint\frac{\udl{z_1}}{\prod_{a=1}^4\thetafunc{s\,u_az^{\pm1}}}=\frac{\thetafunc{q\,s^{-4}}}{\prod_{a<b}^4\thetafunc{s^2u_au_b}}=\mathcal{I}_{\mathcal{T}_{\text{B}}}\,,
\label{idN1}
\ee
where we defined the following integration measure over $USp(2N)$ gauge fugacities
\be
\udl{\vec{z}_N}=\frac{\qfac{q}^{2N}}{2^N N!}\prod_{i=1}^N\frac{\udl{z_i}}{2\pi iz_i}\thetafunc{z_i^{\pm2}}\prod_{i<j}^N\thetafunc{z_i^{\pm1}z_j^{\pm1}}
\ee
and we introduced fugacities $u_a$, $s$ in the Cartan of the global symmetry group $SU(4)_u\times U(1)_s$, with the constraint $\prod_{a=1}^4u_a=1$. The integral on the l.h.s.~of \eqref{idN1} is defined with the prescription of taking the residues only at poles coming from fields with the same charge under the $U(1)$ in the Cartan of the $SU(2)$ gauge symmetry. This integral identity first appeared in \cite{Putrov:2015jpa} and one way to test it consists of expanding perturbatively in $q$ both sides and matching them order by order.

In Section \ref{secondgen2d} we will use a slightly different version of the duality. This is obtained adding on both sides of the duality two Fermi fields, which are singlets under the gauge symmetry of theory $\mathcal{T}_{\text{A}}$, and coupling them to some of the gauge invariant operators. Specifically, on the side of theory $\mathcal{T}_{\text{A}}$ we split the 4 chirals $Q_a$ into two pairs of chirals $L_i$, $R_i$ with $i=1,2$ and we add two Fermi singlets $\Psi_L$, $\Psi_R$ that flip\footnote{The procedure of ``flipping" an operator $\mathcal{O}$ consists of adding a gauge singlet $S$ together with the superpotential deformation $\gd\mathcal{W}=S\,\mathcal{O}$, so that the equation of motion of $S$ sets $\mathcal{O}$ to zero. In order to preserve the duality, one should consistently add the same singlets to both of the theories.} the mesonic operators
\be
W_{\mathcal{T}_{\text{A}}}=\Psi_LL\,L+\Psi_RR\,R\,.
\ee
This has the effect of explicitly breaking the original $SU(4)_u$ global symmetry to the subgroup $SU(2)_l\times SU(2)_r\times U(1)_d$. Schematically, we can represent the new version of theory $\mathcal{T}_{\text{A}}$ with the following quiver diagram:
\begin{figure}[h]
	\centering
	\makebox[\linewidth][c]{
  	\includegraphics[scale=0.55]{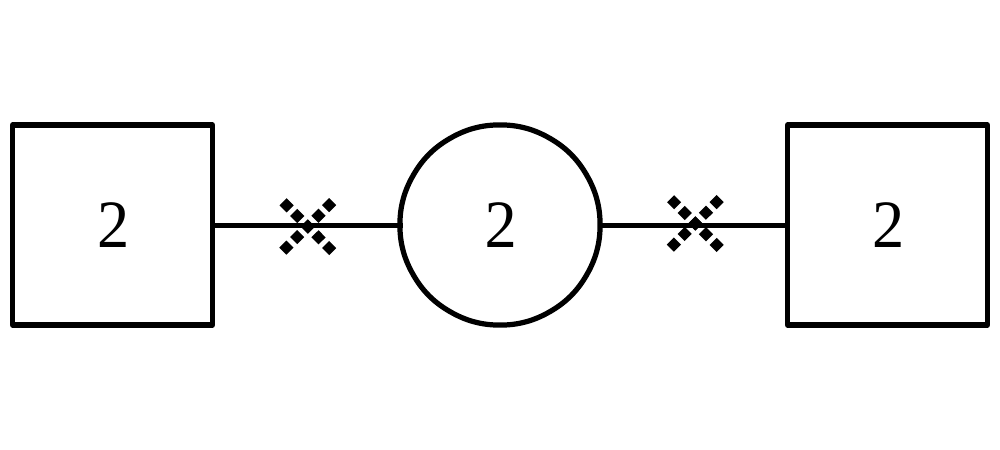} }
\end{figure}

\noindent where the circle node with the label $2$ denotes the $SU(2)$ gauge symmetry, the two square nodes with the label 2 denote the $SU(2)_l\times SU(2)_r$ global symmetries, the straight lines represent the chiral multiplets $L$, $R$ and the two dashed crosses represent the Fermi singlets $\Psi_L$, $\Psi_R$.

On the side of theory $\mathcal{T}_{\text{B}}$ the deformation has the effect of making both the new Fermi fields $\Psi_L$, $\Psi_R$ and two of the original six chirals $\Gp_{ab}$, specifically those uncharged under the $SU(2)_l\times SU(2)_r$ subgroup of $SU(4)_u$, massive. Hence, we end up with an $SU(2)_l\times SU(2)_r$ bifundamental chiral field $Q_{ij}$ and a Fermi field $\Psi$ interacting with\footnote{We will often omit contractions of gauge and flavor indices, which should be understood from the context. For example in this case $\Psi \,Q\,Q=\epsilon^{ij}\epsilon^{kl}\Psi\,Q_{ik}Q_{jl}$.}
\be
W_{\mathcal{T}_{\text{B}}}=\Psi \,Q\,Q\,.
\ee
Schematically, we can represent the new version of theory $\mathcal{T}_{\text{B}}$ with the following quiver diagram:
\begin{figure}[h]
	\centering
	\makebox[\linewidth][c]{
  	\includegraphics[scale=0.55]{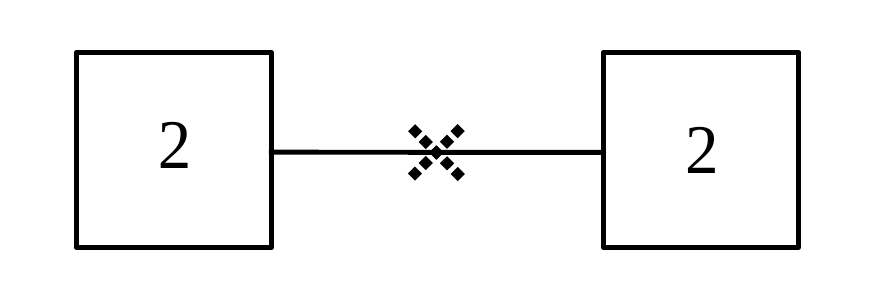} }
\end{figure}

\newpage
The equality of the elliptic genera associated to this duality, which we will use intensively in Subsubsection \ref{EGgen2}, is obtained from \eqref{idN1} by simply re-expressing it in  terms of the fugacities for the subgroup $SU(2)_l\times SU(2)_r\times U(1)_d\subset SU(4)_u$ and moving the contributions of two of the chiral fields from the r.h.s.~to the l.h.s.~\footnote{We also used the property of the theta-function $\thetafunc{x}=\thetafunc{q\,x^{-1}}$ which trivially follows from its definition $\thetafunc{x}=\qfac{x}\qfac{q\,x^{-1}}$, where $\qfac{x}=\prod_{k=0}^\infty(1-x\, q^k)$. This is the translation at the level of the elliptic genus of the fact that a Fermi multiplet in a representation $\mathcal{R}$ is equivalent to a Fermi multiplet in a representation $\bar{\mathcal{R}}$ with $E$ and $J$ interactions swapped.}
\be
\mathcal{I}_{\mathcal{T}_{\text{A}}}=\thetafunc{q\,s^{-2}d^{\pm2}}\oint\frac{\udl{z_1}}{\thetafunc{s\,d\,l^{\pm1}z^{\pm1}}\thetafunc{s\,d^{-1}r^{\pm1}z^{\pm1}}}=\frac{\thetafunc{q\,s^{-4}}}{\thetafunc{s^2l^{\pm1}r^{\pm1}}}=\mathcal{I}_{\mathcal{T}_{\text{B}}}\,.\nn\\
\label{idN1alt}
\ee
Observe that the fugacity $d$ completely disappeared from the expression for $\mathcal{I}_{\mathcal{T}_{\text{B}}}$. Indeed, $U(1)_d$ is not a symmetry of this alternative version of theory $\mathcal{T}_{\text{B}}$, since no fields charged under it remained. The equality \eqref{idN1alt} then implies that even if we refine the elliptic genus of theory $\mathcal{T}_{\text{A}}$ with the fugacity for $U(1)_d$, in the end it is actually independent of $d$. This fact can be checked by computing perturbatively in $q$ the elliptic genus $\mathcal{I}_{\mathcal{T}_{\text{A}}}$. This means that $U(1)_d$ is not actually a symmetry of the low energy theory to which $\mathcal{T}_{\text{A}}$ flows. We will come back to this point in Subsection \ref{secondgen2d}.

\subsection{Dimensional reduction of $4d$ $\mathcal{N}=1$ dualities on $\mathbb{S}^2$}
\label{dimredgen}

The dimensional reduction of $4d$ $\mathcal{N}=1$ theories to $2d$ $\mathcal{N}=(0,2)$ theories has been discussed in details in \cite{Gadde:2015wta}. The first step consists of defining the four-dimensional theory on $\mathbb{R}^2\times\mathbb{S}^2$ preserving half of its supersymmetry. As discussed also in \cite{Closset:2013sxa,Benini:2015noa,Honda:2015yha}, this can be done by introducing a background vector multiplet for a $U(1)_R^{\text{4d}}$ R-symmetry with a quantized flux through $\mathbb{S}^2$ so to cancel the contribution of the spin connection of $\mathbb{S}^2$ in the supersymmetry variation of the fermionic fields. In this way we preserve two supercharges with same chirality on $\mathbb{R}^2$, giving a $2d$ theory with $\mathcal{N}=(0,2)$ supersymmetry. The quantization of the flux translates into the requirement  that the $U(1)_R^{\text{4d}}$ symmetry should be such that all the chiral fields of the theory have integer R-charge. This $U(1)_R^{\text{4d}}$ doesn't have to be the superconformal one of the theory to which our $4d$ $\mathcal{N}=1$ theory flows in the IR, but it has to be non-anomalous. Indeed, it is usually chosen taking specific mixing coefficients with any $U(1)$ in the Cartan of the global symmetries of the theory, possibly also non-abelian ones, which don't necessarily correspond to the superconformal R-symmetry. This procedure is also known as \emph{topological twist} \cite{Witten:1988ze,Witten:1991zz}.

When defined on $\mathbb{R}^2\times\mathbb{S}^2$, each multiplet of the $4d$ $\mathcal{N}=1$ theory decomposes in an infinite tower of KK modes which can be re-arranged into different $2d$ $\mathcal{N}=(0,2)$ multiplets \cite{Kutasov:2013ffl}. This can also be understood from the $\mathbb{T}^2\times\mathbb{S}^2$ partition function of the $4d$ $\mathcal{N}=1$ theory, which was computed using localization methods in \cite{Benini:2015noa} provided that we choose an integrally quantized R-symmetry $U(1)_R^{\text{4d}}$ as we just discussed. Such a partition function takes indeed the form of an infinite sum of contour integrals that are precisely of the form of elliptic genera of $2d$ $\mathcal{N}=(0,2)$ theories. In \cite{Gadde:2015wta} it was shown that this sum actually truncates and that it reduces to a single term corresponding to the zero modes of the $4d$ fields provided that we also choose $U(1)_R^{\text{4d}}$ to be non-negative\footnote{The dimensional reduction from $4d$ $\mathcal{N}=1$ to $2d$ $\mathcal{N}=(0,2)$ without this constraint on the non-negativity of the R-charges was first discussed in \cite{tachi}.}. Hence, with this choice we get a single $2d$ $\mathcal{N}=(0,2)$ theory, which is the one describing the zero modes of the $4d$ fields.

The zero modes of the $4d$ fields can be re-organized into $2d$ $\mathcal{N}=(0,2)$ multiplets according to the following rules \cite{Gadde:2015wta,Kutasov:2013ffl}. A $4d$ $\mathcal{N}=1$ chiral multiplet of R-charge $r$ gives
\begin{itemize}
\item $r-1$ Fermi multiplets when $r>1$;
\item $1-r$ chiral multiplets when $r<1$;
\item no multiplets when $r=1$.
\end{itemize}
Instead, the zero modes of a $4d$ $\mathcal{N}=1$ vector multiplet consist only of a $2d$ $\mathcal{N}=(0,2)$ vector multiplet. Hence, a four-dimensional gauge theory will reduce to a two-dimensional gauge theory with the same gauge group, but with different matter content. These matter fields will have interactions that can be understood from the original superpotential of the $4d$ theory.

If we apply this procedure to two dual $4d$ $\mathcal{N}=1$ theories, we obtain a putative $2d$ $\mathcal{N}=(0,2)$ duality. As we mentioned in the introduction, it is not always true that the resulting duality is valid and one should perform all the standard tests to understand if this is the case or not. Moreover, there are in principle many choices of the R-symmetry $U(1)_R^{\text{4d}}$ that we can make, which lead to different possible $2d$ dualities starting from a single duality in $4d$. Nevertheless, these possibilities are strictly constrained from the requirement that the $4d$ R-charges should be non-negative integers and from the cancellation of gauge anomalies in the resulting $2d$ theories.

As a particular example of the application of the prescription of \cite{Gadde:2015wta} we will review how to obtain the duality discussed in the previous subsection from Seiberg duality in four dimensions \cite{Seiberg:1994pq}. More precisely, the $4d$ $\mathcal{N}=1$ duality we should start with is the one relating the $SU(2)$ gauge theory with 6 fundamental chirals and the Wess--Zumino (WZ) model of 15 chirals with a cubic superpotential, which corresponds to the case $N_c=2$ and $N_f=3$ of Seiberg duality. 

The global symmetry of the $4d$ gauge theory is $SU(6)_v$, since the $U(1)$ part is anomalous. In particular, requiring the existence of a non-anomalous R-symmetry we can uniquely fix the superconformal R-charge of the chirals, which has to be $\frac{1}{3}$. Nevertheless, we know that in the dimensional reduction we can choose another R-symmetry $U(1)_R^{\text{4d}}$ that differs for a mixing with any $U(1)$ in the Cartan of the $SU(6)_v$ flavor symmetry. In order to choose the correct $U(1)_R^{\text{4d}}$ symmetry that reproduces the $2d$ duality, we consider the subgroup $SU(4)_u\times SU(2)\times U(1)_s\subset SU(6)_v$ of the global symmetry\footnote{This is not the only choice leading to a consistent two-dimensional duality. For other choices of $U(1)_R^{\text{4d}}$ see \cite{Gadde:2015wta}.}. We then allow for a mixing of the non-anomalous R-symmetry with $U(1)_s$. The chirals which transform in the fundamental representation of $SU(6)_v$ accordingly decompose as
\be
{\bf 6}\rightarrow({\bf 4},{\bf 1})^1\oplus({\bf 1},{\bf 2})^{-2}\,.
\label{BRfundSU6}
\ee
Choosing the mixing coefficient with $U(1)_s$ to be $R_s=-\frac{1}{3}$ we see that the four chirals in the representation $({\bf 4},{\bf 1})^1$ have R-charge $\frac{1}{3}+R_s=0$ and become four chirals in the $2d$ theory, while the two chirals in the representation $({\bf 1},{\bf 2})^{-2}$ have R-charge $\frac{1}{3}-2R_s=1$ and don't survive the dimensional reduction. In conclusion, on this side of the duality we get a $2d$ $\mathcal{N}=(0,2)$ $SU(2)$ gauge theory with 4 fundamental chirals.

On the other side of the duality we have 15 chirals that can be collected in a matrix $M$ transforming in the antisymmetric representation of $SU(6)_v$ and which interact with cubic superpotential
\be
\mathcal{W}=\mathrm{Pf}\,M\,.
\ee
This is a cubic interaction that fixes the R-charges of all the chirals to $\frac{2}{3}$. In this case we use the branching rule
\be
{\bf 15}\rightarrow({\bf 4},{\bf 2})^{-1}\oplus({\bf 6},{\bf 1})^{2}\oplus({\bf 1},{\bf 1})^{-4}\,.
\label{BRantisymmSU6}
\ee
and choosing $R_s=-\frac{1}{3}$ we see that only the representations $({\bf 6},{\bf 1})^{2}$ and $({\bf 1},{\bf 1})^{-4}$ survive the dimensional reduction, becoming the chirals $\Gp_{ab}$ and the Fermi $\Psi$ respectively. Of the original $4d$ superpotential, we are left with the $J$-interaction
\be
W_{\mathcal{T}_{\text{B}}}=\Psi\,\mathrm{Pf}\,\Gp\,.
\ee
We thus recovered the duality discussed in the previous subsection. Notice also that of the original $4d$ global symmetry $SU(4)_u\times SU(2)\times U(1)_s$ we are left only with $SU(4)_u\times U(1)_s$, since all the fields charged under the $SU(2)$ factor didn't survive the dimensional reduction.

\subsection{$2d$ $\mathcal{N}=(0,2)$ version of the confining Intriligator--Pouliot duality}
\label{dimredIP}

We conclude this review part with the dimensional reduction of the four-dimensional Intriligator--Pouliot duality \cite{Intriligator:1995ne}, which has been discussed in \cite{Gadde:2015wta}. We will focus on the confining case $N_c=N$, $N_f=N+2$ since for different values of $N_f$ we don't get new $2d$ dualities. This is the following IR duality between $4d$ $\mathcal{N}=1$ theories:

\medskip
\noindent \textbf{Theory \boldmath$\mathcal{T}_A^{\text{4d}}$}: $USp(2N)$ gauge theory with $2N+4$ fundamental chirals and no superpotential
\be
\mathcal{W}_{\mathcal{T}_A^{\text{4d}}}=0\,.
\ee

\medskip
\noindent \textbf{Theory \boldmath$\mathcal{T}_B^{\text{4d}}$}: WZ model of $(N+2)(2N+3)$ chirals collected in an antisymmetric $(2N+4)\times(2N+4)$ matrix $M_{ab}$ for $a<b=1,\cdots,2N+4$ interacting with a  superpotential of degree $N+2$
\be 
\mathcal{W}_{\mathcal{T}_B^{\text{4d}}}=\mathrm{Pf}\,M\,.
\ee

Notice that for $N=1$ this reduces to the Seiberg duality we considered in the last subsection.
The non-anomalous global symmetry of the dual theories is $SU(2N+4)_v$. 
Again the requirement of the existence of a non-anomalous R-symmetry fixes the R-charges of the chirals in the gauge theory to be $\frac{1}{N+2}$, while on the WZ dual side the R-charges of the chirals $M$ are fixed by the superpotential to the value $\frac{2}{N+2}$. The R-symmetry $U(1)_R^{\text{4d}}$ we want to use in the reduction is obtained decomposing $SU(2N+2)_u\times SU(2)\times U(1)_s\subset SU(2N+4)_v$ and considering a mixing $R_s$ of the non-anomalous R-symmetry we just found with $U(1)_s$. Under this subgroup, the fundamental and the antisymmetric representations of $SU(2N+4)_v$ decompose according to
\be
{\bf 2N+4}&\rightarrow&({\bf 2N+2},{\bf 1})^1\oplus({\bf 1},{\bf 2})^{-(N+1)}\nn\\
{\bf (N+2)(2N+3)}&\rightarrow&({\bf 2N+2},{\bf 2})^{-N}\oplus({\bf (N+1)(2N+1)},{\bf 1})^{2}\oplus({\bf 1},{\bf 2})^{-2(N+1)}\,.
\ee
Choosing $R_s=-\frac{1}{N+2}$ we get the following $2d$ $\mathcal{N}=(0,2)$ multiplets:
\begin{itemize}
\item on the side of theory $\mathcal{T}_A^{\text{4d}}$ we have $2N+2$ chiral multiplets forming the fundamental representation of $SU(2N+2)_u$ which have charge 1 under $U(1)_s$;
\item on the side of theory $\mathcal{T}_B^{\text{4d}}$ we have $(N+1)(2N+1)$ chiral multiplets forming the antisymmetric representation of $SU(2N+2)_u$ which have charge 2 under $U(1)_s$ and one Fermi multiplet which is a singlet of $SU(2N+2)_u$ and has charge $-2(N+1)$ under $U(1)_s$.
\end{itemize}

The $2d$ $\mathcal{N}=(0,2)$ duality we get from the dimensional reduction of the $4d$ $\mathcal{N}=1$ Intriligator--Pouliot duality in the confining case is

\medskip
\noindent \textbf{Theory \boldmath$\mathcal{T}_{\text{A}}$}: $USp(2N)$ gauge theory with $2N+2$ fundamental chirals and no superpotential
\be
W_{\mathcal{T}_{\text{A}}}=0\,.
\ee

\medskip
\noindent \textbf{Theory \boldmath$\mathcal{T}_{\text{B}}$}: LG model of one Fermi $\Psi$ and $(N+1)(2N+1)$ chirals collected in an antisymmetric $(2N+2)\times(2N+2)$ matrix $\Gp_{ab}$ for $a<b=1,\cdots,2N+2$ interacting with a superpotential of degree $N+2$
\be 
W_{\mathcal{T}_{\text{B}}}=\Psi\,\mathrm{Pf}\,\Gp\,.
\ee

Notice that for $N=1$ this reduces to the duality we reviewed in Subsection \ref{review1}. The transformation rules of the fields of the two theories under the $SU(2N+2)_u\times U(1)_s$ global symmetry, which can either be obtained from the $4d$ ones or from the superpotential constraints, are summarized in the following table:
\begin{table}[h]
\centering
\scalebox{1}{
\setlength{\extrarowheight}{1pt}
\begin{tabular}{c|cc|c}
{} & $SU(2N+2)$ & $U(1)_s$ & $U(1)_{R_0}$ \\ \hline
$Q$ & $\bf 2N+2$ & 1 & $0$ \\ \hline
$\Psi$ & $\bullet$ & $-2(N+1)$ & $1$ \\
$\Gp$ & $\bf (N+1)(2N+1)$ & 2 & $0$
\end{tabular}}
\end{table}

One simple test we can perform for this duality consists of matching anomalies
\be
\Tr\,\gc^3\,U(1)_s^2=4N(N+1), \qquad \Tr\,\gc^3\,SU(2N+2)_u^2=N\,.
\label{anomaliesconfiningIP2d}
\ee
We can also compute the trial central charges
\be
c_R=3 N \left(4 (N+1) (R_s-1)^2-2 N-1\right),\qquad c_R-c_L=N(2N+3)\,.
\ee
where again $U(1)_R$ is defined taking into account a generic mixing with $U(1)_s$ as in \eqref{trialRsymmetry}. If we naively try to perform $c$-extremization we get, as in the $N=1$ case, $R_s=1$ and a corresponding value of the central charges
\be
c_R^{\text{naive}}=-3N(2N+1),\qquad c_L^{\text{naive}}=-2N(4N+3)\,.
\ee
This result can't be correct, as the central charges are negative for any $N$. Again we interpret this as due to the fact that the theories have a non-compact target space, since the number of chirals on the gauge theory side is large enough to expect no SUSY breaking in the IR (see footnote 6). From the analysis of the equations of motion, we can easily identify the classical moduli space on the side of theory $\mathcal{T}_{\text{B}}$. This is the hypersurface in $\mathbb{C}^{(N+1)(2N+1)}$ defined by the polynomial equation
\be
\mathrm{Pf}\,\Gp=0\,.
\ee
Similarly to the $N=1$ case, we have to require that the chirals $\Gp_{ab}$ parametrizing this non-compact target space have vainishing R-charge. In this way, we fix the mixing coefficient $R_s=0$ and, consequently, the central charges
\be
c_R=3N(2N+3),\qquad c_L=2N(2N+3)\,.
\ee

Another test is matching the elliptic genera of the two theories. In particular, the duality translates into the following integral identity
\be
\mathcal{I}_{\mathcal{T}_{\text{A}}}=\oint\frac{\udl{\vec{z}_N}}{\prod_{i=1}^N\prod_{a=1}^{2N+2}\thetafunc{s\,u_az^{\pm1}_i}}=\frac{\thetafunc{q\,s^{-2(N+1)}}}{\prod_{a<b}^{2N+2}\thetafunc{s^2u_au_b}}=\mathcal{I}_{\mathcal{T}_{\text{B}}}\,,
\label{idIP2d}
\ee
This equality can also be understood as the matching of the $\mathbb{T}^2\times\mathbb{S}^2$ partition functions of the original $4d$ dual theories. We couldn't find an analytical proof of this result in the mathematical literature, as instead can be done for the matching of the $\mathbb{S}^3\times \mathbb{S}^1$ partition functions of the $4d$ theories \cite{10.1155/S1073792801000526,2003math......9252R}. Nevertheless, this identity can be tested perturbatively in $q$ for low values of the rank $N$. As we will show in Subsubsection \ref{EGsecgen}, equation \eqref{idIP2d} will play a key role in the derivation of the identity for one of the dualities we are going to propose. 

\section{Duality for $USp(2N)$ gauge theory with antisymmetric matter}
\label{sec3}

In this section we discuss an higher rank generalization of the duality for the $SU(2)$ theory with 4 fundamental chirals, where instead of increasing the number of fundamental flavors as in the Intriligator--Pouliot duality we add a chiral field in the antisymmetric representation of the gauge group.

If we consider a $2d$ $\mathcal{N}=(0,2)$ theory with $USp(2N)$ gauge group, one antisymmetric chiral, $N_b$ fundamental chirals and $N_f$ fundamental Fermis, cancellation of gauge anomalies requires that $N_b-N_f-4=0$. We would like to find a dual for this class of theories. We will do so starting from a $4d$ duality for a theory with 6 fundamental chirals only. This means that from it we can only derive a $2d$ duality for a theory with $N_b+N_f\le 6$. Combining this constraint with the one coming from anomaly cancellation we see that we can have only two possibilities: $N_b=4$, $N_f=0$ or $N_b=5$, $N_f=1$. In this section we will discuss in detail the former case, while we will comment on the latter in Appendix \ref{appA}.

We start reviewing the $4d$ ancestor duality, then we discuss its dimensional reduction and finally we perform some tests for the resulting $2d$ duality. In particular, we show how to derive the proposed $2d$ duality by iterative applications of the duality we reviewed in Subsection \ref{dimredIP} corresponding to the dimensional reduction of Intriligator--Pouliot duality in the confining case, in complete analogy to a similar derivation that can be done for the original $4d$ duality.

\subsection{The $4d$ duality}
\label{firstgen4d}

The $4d$ duality we are interested in was first proposed by Csaki, Skiba and  Schmaltz in \cite{Csaki:1996eu}:

\medskip
\noindent \textbf{Theory \boldmath$\mathcal{T}_A^{\text{4d}}$}: $USp(2N)$ gauge theory with one antisymmetric chiral $A$, six fundamental chirals $Q_a$ and $N$ chiral singlets $\beta_i$ with superpotential\footnote{Throughout the paper, all the $USp(2n)$ indices are contracted using the totally antisymmetric tensor
\be
J^{(n)}=\mathbb{I}_n\otimes i\,\gs_2\, .
\ee
For example, the trace of the $USp(2n)$ antisymmetric operator $A$ is
\be
\Tr_n\,A=J^{(n)}_{ij}A^{ij}\,.
\ee}
\be
\mathcal{W}_{\mathcal{T}_A^{\text{4d}}}=\sum_{i=1}^N\gb_i \Tr_NA^i\, .
\ee

\medskip
\noindent \textbf{Theory \boldmath$\mathcal{T}_B^{\text{4d}}$}: WZ model with $15N$ chiral singlets $\mu_{ab;i}$ for $i=1,\cdots,N$, $a<b=1,\cdots,6$ interacting with the cubic superpotential
\be
\mathcal{W}_{\mathcal{T}_B^{\text{4d}}}=\sum_{i,j,k=1}^N\sum_{a,b,c,d,e,f=1}^6\epsilon_{abcdef}\mu_{ab;i}\mu_{cd;j}\mu_{ef;k}\gd_{i+j+k,2N+1}\, .
\ee

Notice that for $N=1$ this duality reduces to the Seiberg duality between $SU(2)$ with 6 chirals and the WZ model of 15 chirals whose dimensional reduction we studied in Subsection \ref{dimredgen}. Indeed, the antisymmetric of $SU(2)$ is just a singlet and the superpotential $\mathcal{W}=\gb_1\,A$ is a mass term for both the singlets $\gb_1$ and $A$. Integrating them out we recover the aforementioned duality.

The non-anomalous global symmetry of the dual theories is
\be
SU(6)_v\times U(1)_x\,,
\ee
under which the chiral fields transform according to
\begin{table}[h]
\centering
\scalebox{1}{\setlength{\extrarowheight}{2pt}
\begin{tabular}{c|cc|c}
{} & $SU(6)_v$ & $U(1)_x$ & $U(1)_{R_0}$\\ \hline
$\gb_i$ & $\bullet$ & $-i$ & 2 \\
$Q$ & $\bf 6$ & $\frac{1-N}{3}$ & $\frac{1}{3}$ \\
$A$ &  $\bullet$ & 1 & 0 \\ \hline
$\mu_{i}$ & $\bf 15$ & $i-\frac{2N+1}{3}$ & $\frac{2}{3}$
\end{tabular}}
\end{table}

\noindent where $U(1)_{R_0}$ is a possible choice of UV trial R-symmetry. The charges under $U(1)_x$ are determined on the gauge theory side requiring that $U(1)_R$ is not anomalous, where
\be
R=R_0+q_x\,R_x\,,
\ee
with $q_x$ being the charge under $U(1)_x$ and $R_x$ the mixing coefficient of $U(1)_x$ with the R-symmetry, while on the WZ side they are fixed by the superpotential.

This duality can be derived by iterative applications of the Intriligator--Pouliot duality in the confining case. This strategy was used in \cite{2003math......9252R} to prove the equality of the $\mathbb{S}^3\times \mathbb{S}^1$ partition functions of the dual theories. In Subsubsection  \ref{EGsecgen} we will present a similar derivation for the $2d$ version of the duality that we are going to discuss. 

\subsection{Dimensional reduction}
\label{firstgendimred}

In order to dimensionally reduce the $4d$ duality, we follow a strategy similar to the one we used in Subsection \ref{dimredgen}. We first decompose the non-abelian part of the global symmetry into $SU(4)_u\times SU(2)\times U(1)_s\subset SU(6)_v$\footnote{In Appendix \ref{appA} we will consider the only other possible decomposition that leads to a potential $2d$ duality, which will be for a theory with $N_b=5$ fundamental chirals and $N_f=1$ fundamental Fermis.}. Then we introduce a possible mixing of $U(1)_x$ and $U(1)_s$ with the trial R-symmetry $U(1)_{R_0}$
\be
R=R_0+q_x\,R_x+q_s\,R_s\,.
\ee
The charges of the chiral fields of theory $\mathcal{T}_A^{\text{4d}}$ under this R-symmetry are
\be
&&R[Q_\ga]=\frac{1}{3}+\frac{1-N}{3}R_x-2R_s,\qquad \ga=1,2\nn\\
&&R[Q_a]=\frac{1}{3}+\frac{1-N}{3}R_x+R_s,\qquad a=3,\cdots,6\nn\\
&&R[A]=R_x\nn\\
&&R[\gb_i]=2-i\,R_x,\qquad i=1,\cdots, N
\ee
where we used the charges under $U(1)_s$ dictated by the branching rule \eqref{BRfundSU6}. For theory $\mathcal{T}_B^{\text{4d}}$ we have
\be
&&R[\mu_{12;i}]=\frac{2}{3}+\left(i-\frac{2N+1}{3}R_x\right)-4R_s\nn\\
&&R[\mu_{ab;i}]=\frac{2}{3}+\left(i-\frac{2N+1}{3}R_x\right)+2R_s,\qquad a<b=3,\cdots,6\nn\\
&&R[\mu_{\ga a;i}]=\frac{2}{3}+\left(i-\frac{2N+1}{3}R_x\right)-R_s,\qquad\ga=1,2, \quad a=3,\cdots, 6\,,
\ee
where we instead used the branching rule \eqref{BRantisymmSU6}. The R-symmetry $U(1)_R^{\text{4d}}$ we choose for the dimensional reduction corresponds to $R_x=0$ and $R_s=-\frac{1}{3}$. With this choice we get the following $2d$ $\mathcal{N}=(0,2)$ multiplets:
\begin{itemize}
\item on the side of theory $\mathcal{T}_A^{\text{4d}}$ the fields $Q_\ga$ don't survive the dimensional reduction, while $Q_a$, $A$ become chirals and $\gb_i$ become Fermi fields;
\item on the side of theory $\mathcal{T}_B^{\text{4d}}$ the fields $\mu_{\ga a;i}$ don't survive the dimensional reduction, while $\mu_{ab;i}$ become chirals and $\mu_{12;i}$ become Fermi fields.
\end{itemize}

\subsection{The $2d$ duality}
\label{firstgen2d}

From the dimensional reduction we got the following putative $2d$ $\mathcal{N}=(0,2)$ duality:

\medskip
\noindent \textbf{Theory \boldmath$\mathcal{T}_{\text{A}}$}: $USp(2N)$ gauge theory with one antisymmetric chiral $A$, four fundamental chirals $Q_a$ and $N$ Fermi singlets $\beta_i$ with superpotential
\be
W_{\mathcal{T}_{\text{A}}}=\sum_{i=1}^N\gb_i \Tr_NA^i\, .
\label{firstgenWA}
\ee

\medskip
\noindent \textbf{Theory \boldmath$\mathcal{T}_{\text{B}}$}: LG model with $N$ Fermi fields $\Psi_i$ and $6N$ chiral fields $\Gp_{ab;i}$ for $i=1,\cdots,N$, $a<b=1,\cdots,4$ interacting with cubic superpotential
\be
W_{\mathcal{T}_{\text{B}}}=\sum_{i,j,k=1}^N\sum_{a,b,c,d=1}^4\epsilon_{abcd}\Psi_i\Gp_{ab;j}\Gp_{cd;k}\gd_{i+j+k,2N+1}\, .
\label{firstgenWB}
\ee

This duality can be understood as a higher rank generalization of the duality for $SU(2)$ with 4 chirals, to which it reduces in the particular case $N=1$. The global symmetry group of the dual theories is 
\be
SU(4)_u\times U(1)_s\times U(1)_x\,,
\ee
where the $U(1)_x$ factor is present only if $N>1$. This symmetry is indeed associated to the antisymmetric chiral $A$, which is a massive singlet together with $\gb_1$ for $N=1$. The fields transform under this symmetry according to
\begin{table}[h]
\centering
\scalebox{1}{
\setlength{\extrarowheight}{1pt}
\begin{tabular}{c|ccc|c}
{} & $SU(4)$ & $U(1)_s$ & $U(1)_x$ & $U(1)_{R_0}$ \\ \hline
$\gb_i$ & $\bullet$ & 0 & $-i$ & 1 \\
$Q$ & $\bf 4$ & 1 & $\frac{1-N}{3}$ & $0$ \\
$A$ & $\bullet$ & 0 & 1 & 0 \\ \hline
$\Psi_i$ & $\bullet$ & $-4$ & $i-\frac{2N+1}{3}$ & 1 \\
$\Gp_i$ & $\bf 6$ & 2 & $i-\frac{2N+1}{3}$ & $0$
\end{tabular}}
\end{table}

\newpage
\noindent The charges under the $U(1)$ symmetries can be determined by imposing superpotential constraint, but they also coincide with those of the original $4d$ fields.

In the following we perform some tests for the validity of this duality.

\subsubsection{Anomalies}
\label{anomaliessec3}

The anomalies for the abelian factors of the global symmetry group are for both of the theories
\be
\Tr\,\gc^3U(1)_s^2=8N,\quad \Tr\,\gc^3U(1)_x^2=\frac{5}{18} N(N-1) (2 N+1),\quad \Tr\,\gc^3U(1)_sU(1)_x=-\frac{8}{3}N(N-1)\,,\nn\\
\ee
while the anomaly for the $SU(4)_u$ factor is
\be
\Tr\,\gc^3SU(4)_u^2=N\,.
\ee
These results coincide with \eqref{anomaliesfundduality} for $N=1$. In particular, the anomalies involving $U(1)_x$ are zero when $N=1$, as expected since this symmetry disappears in this particular case.

We can also match the trial central charges
\be
&&c_R=\frac{1}{6} N \left(-96 R_s ((N-1) R_x+3)+(N-1) R_x (5 (2 N+1) R_x+42)+144 R_s^2+90\right)\nn\\
&&c_R-c_L=5N\,,
\label{trialcentralchargesantisymm}
\ee
which again coincide with \eqref{centrachargesfundduality} for $N=1$, with the mixing coefficient with $U(1)_x$ disappearing. If we try to perform $c$-extremization directly from \eqref{trialcentralchargesantisymm}, we get the naive values of the mixing coefficients
\be
R_s^{\text{naive}}=-\frac{N+4}{2N-7},\qquad R_x^{\text{naive}}=-\frac{9}{2N-7}\,,
\label{mixingcmax}
\ee
from which we find the values of the central charges
\be
c_R^{\text{naive}}=\frac{45N(N+1)}{2(2N-7)}, \qquad c_L^{\text{naive}}=\frac{5N(5N+23)}{2(2N-7)}\,.
\label{centralcmax}
\ee
Notice that these are not positive for any value of the rank $N$ of the gauge group. Once again, this is due to the fact that the target space is non-compact, as it can be understood for example from the computation of the elliptic genus (see footnote 18 in the next subsubsection). Moreover, we can see that the moduli space is non-compact already at the classical level. Indeed, on the side of theory $\mathcal{T}_{\text{B}}$ the equations of motion of the Fermi fields $\Psi_i$ imply that the classical moduli space is a co-dimension $N$ subspace of $\mathbb{C}^{6N}$ defined by the following $N$ independent polynomial equations:
\be
\sum_{j=0}^{N-1}\epsilon_{abcd}\Gp_{ab;N-j}\Gp_{cd;N-i+j+1}=0,\qquad i=1,\cdots,N\,,
\ee
where we defined $\Gp_i=0$ for all $i>N$. Notice that in particular for $N=1$ we get $\mathrm{Pf}\,\Gp_1=\epsilon_{abcd}\Gp_{ab;1}\Gp_{cd;1}=0$ which corresponds to \eqref{eom}. Hence, in order to determine the correct conformal central charges we have to require that the R-charges of the chirals $\Gp_{ab;i}$ are zero for every $i=1,\cdots,N$ and for every $a<b=1,\cdots,4$. This fixes the two mixing coefficients to the values
\be
R_s=R_x=0
\ee
and, consequently, the central charges to
\be
c_R=15N,\qquad c_L=10N\,,
\ee
which are now positive for any $N$.

\subsubsection{Elliptic genus and derivation}
\label{EGsecgen}

In this section we show how to derive the proposed duality by iterative applications of the $2d$ version of the confining Intriligator--Pouliot duality we reviewed in Subsection \ref{dimredIP}. We will first sketch the derivation at the level of quivers and then apply it in details to derive the corresponding equality of elliptic genera.

We remark that a completely analogous derivation exists for the original $4d$ duality of \cite{Csaki:1996eu}, which is based on iterative applications of the confining Intriligator--Pouliot duality \cite{Intriligator:1995ne}. This strategy was used in \cite{2003math......9252R} to prove the equality of the $\mathbb{S}^3\times\mathbb{S}^1$ partition functions of the dual theories. Moreover, it also exists a $3d$ $\mathcal{N}=2$ version of the duality, which is obtained from the $4d$ one by compactification on $\mathbb{S}^1$ followed by a series of real mass deformations \cite{Benvenuti:2018bav}. This three-dimensional duality relates a $U(N)$ gauge theory with one adjoint chiral $A$, one fundamental flavor $Q$ and $N$ chiral singlets with superpotential $\mathcal{W}=\sum_{i=1}^N\gb_i\Tr\,A^i$ and a WZ model of $3N$ chirals $\ga_i$, $T_i^\pm$ interacting with the cubic superpotential $\hat{\mathcal{W}}=\sum_{i,j,l=1}^N\ga_iT_j^+T_{N-l+1}^-\gd_{i+j+l,2N+1}$\footnote{In \cite{Benvenuti:2018bav,Amariti:2018wht} it was actually proposed another $3d$ $\mathcal{N}=2$ duality that is more similar to the $2d$ one we are discussing here, which relates a $USp(2N)$ gauge theory with one antisymmetric chiral and 4 fundamental chirals to a WZ model of $7N$ chirals interacting with a cubic superpotential. We expect that also this duality can be derived by iterative applications of some dimensional reduction to $3d$ of Intriligator--Pouliot duality.}. Also this duality can be derived by iterative applications of some more fundamental confining dualities, as it was shown in \cite{Pasquetti:2019uop} at the level of the $\mathbb{S}^3_b$ partition function\footnote{In \cite{Pasquetti:2019tix} a similar strategy was used to derive a generalization of the $3d$ duality to an higher number of fundamental flavors. It would be interesting to find a $4d$ version of such $3d$ duality that generalizes the one of \cite{Csaki:1996eu} we are considering here and study a possible $2d$ $\mathcal{N}=(0,2)$ reduction \cite{4drankstab}.}. Such more fundamental dualities correspond to the confining cases of Aharony duality \cite{Aharony:1997gp} and a variant with monopole superpotential \cite{Benini:2017dud}, which can also be derived as limits of the compactification of the $4d$ Intriligator--Pouliot duality. Hence, the $2d$ duality we are proposing and its derivation complete this picture: we have an analogue of the same duality in $2d$, $3d$ and $4d$, as well as an analogue of their derivation by iteration of confining dualities.

The derivation consists of two fundamental steps that are iterated $N$ times. At the first step, we start with the $USp(2N)$ gauge theory and trade the antisymmetric chiral field $A$ for an auxiliary $USp(2(N-1))$ gauge node by using the $2d$ version of Intriligator--Pouliot duality we reviewed in Subsection \ref{dimredIP}. In the process one of the Fermi singlets, specifically $\gb_N$, becomes massive so that we get the following dual frame:

\vspace{0.36cm}
\hspace{2.2cm}{\Large $\mathcal{T}_{\text{A}}$ \hspace{7.3cm} $\mathcal{T}_{\text{aux}}^{(1)}$}
\vspace{-0.4cm}
\begin{figure}[h]
	\centering
	\makebox[\linewidth][c]{
  	\includegraphics[scale=0.55]{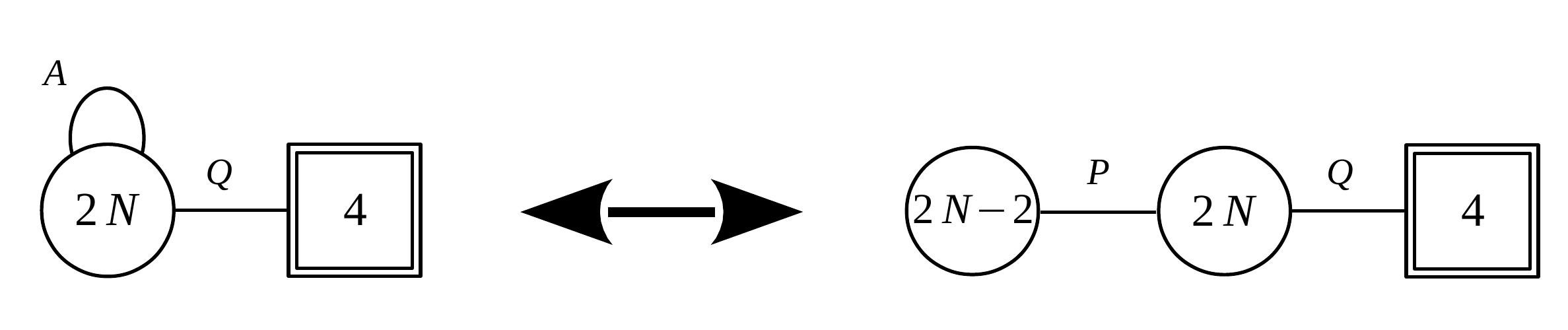} }
\end{figure}

\vspace{-0.32cm}
\hspace{0.7cm}$\mathcal{W}_{\mathcal{T}_{\text{A}}}=\sum_{i=1}^N\gb_i\Tr_NA^i$ \hspace{3.2cm} $\mathcal{W}_{\mathcal{T}_{\text{aux}}^{(1)}}=\sum_{i=1}^{N-1}\gb_i\Tr_N\left(\Tr_{N-1}P\,P\right)^i$
\vspace{0.6cm}

\noindent
In this quiver notation, we represent gauge symmetries with circle nodes and global symmetries with square nodes,  with single lines standing for $USp(2n)$ groups and double lines for $SU(n)$ groups. Solid lines connecting the nodes represent $2d$ $\mathcal{N}=(2,0)$ chiral multiplets charged under the corresponding symmetries.

At this point we observe that in the new dual frame $\mathcal{T}_{\text{aux}}^{(1)}$ we found,  the fields charged under the original $USp(2N)$ gauge node consist of only $2N+2$ fundamental chirals. This means that applying the basic duality again we can confine this node. Recall that the duality produces one Fermi and $(N+1)(2N+1)$ chiral fields. The Fermi singlet will correspond to the field $\Psi_N$ of the final theory $\mathcal{T}_{\text{B}}$. Instead, of the $(N+1)(2N+1)$ chiral fields, $(N-1)(2N-3)$ of them become an antisymmetric chiral field of the remaining $USp(2(N-1))$ gauge node, other $8(N-1)$ of them become 4 fundamental chirals of $USp(2(N-1))$ and the remaining $6$ can be identified with the fields $\Gp_{ab;1}$ of the final theory $\mathcal{T}_{\text{B}}$. Hence, we get to the following dual frame:

\vspace{0.35cm}
\hspace{1.6cm}{\Large $\mathcal{T}_{\text{aux}}^{(1)}$ \hspace{7.3cm} $\mathcal{T}_{\text{aux}}^{(2)}$}
\vspace{-0.4cm}
\begin{figure}[h]
	\centering
	\makebox[\linewidth][c]{
  	\includegraphics[scale=0.55]{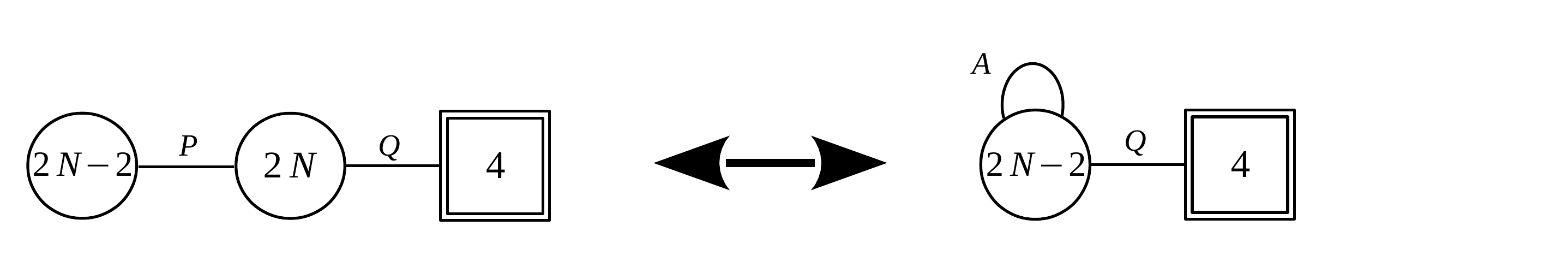} }
\end{figure}

\vspace{-0.32cm}
\hspace{-0.5cm}$\mathcal{W}_{\mathcal{T}_{\text{aux}}^{(1)}}=\sum_{i=1}^{N-1}\gb_i\Tr_N\left(\Tr_{N-1}P\,P\right)^i$ \hspace{3.3cm} $\mathcal{W}_{\mathcal{T}_{\text{aux}}^{(2)}}=\sum_{i=1}^{N-1}\gb_i\Tr_{N-1}A^i+$ 

\vspace{0.1cm}
\hspace{9.1cm}$+\epsilon_{abcd}\Psi_N\Gp_{ab;1}\Tr_{N-1}(Q_cA^{N-2}Q_d)$

\noindent
Comparing this frame $\mathcal{T}_{\text{aux}}^{(2)}$ with the original one of theory $\mathcal{T}_{\text{A}}$, we can see that we recovered the same theory, but with rank of the gauge group decreased by one unit, with one of the $\gb_i$ singlets less and with the addition of one copy of the $\Psi_i$ and $\Gp_{ab;i}$ singlets. 

It is now clear what we should do next: we simply iterate the previous two steps $N$ times, so to completely confine the gauge group, remove all the $\gb_i$ singlets and gain all the $\Psi_i$ and $\Gp_{ab;i}$ singlets
\begin{figure}[h]
	\centering
	\makebox[\linewidth][c]{
  	\includegraphics[scale=0.38]{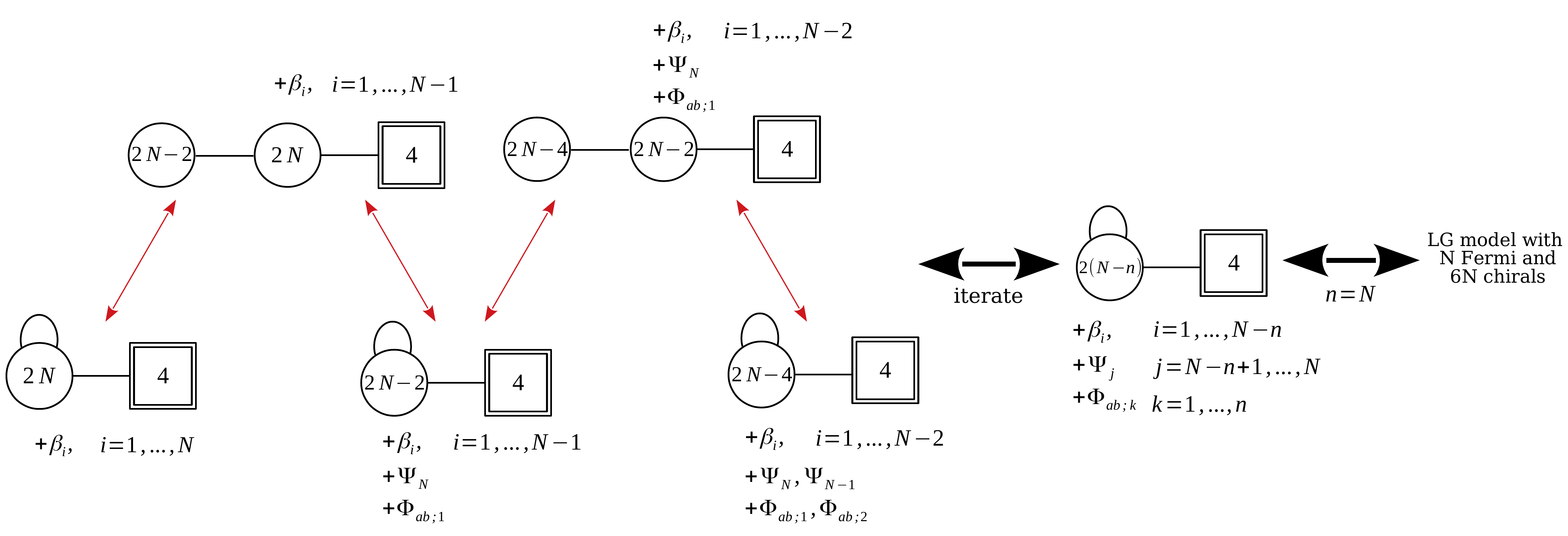} }
\end{figure}

\noindent
The final result is precisely the LG model of $N$ Fermi and $6N$ chirals that we denoted by $\mathcal{T}_{\text{B}}$.

We can use the strategy we just described to prove the equality for the elliptic genera of the two theories. At the level of the elliptic genus, the duality implies the following non-trivial integral identity\footnote{We also verified this identity for $N=2$ with a perturbative computation in $q$. The result explicitly shows that the elliptic genus is non-vanishing, meaning that the theory is not SUSY breaking. Moreover, in the limit in which we turn off the fugacities $u_a\rightarrow1$ the result is divergent, which we interpret as the fact that the
theory has a non-compact Higgs branch (something similar happens, for example, for the $2d$ $\mathcal{N}=(2,2)$ $SU(2)$ gauge theory with $N$ fundamental chirals which has a non-compact Coulomb branch for even $N$ \cite{Hori:2006dk}, as discussed in Sec.~4.4 of \cite{Benini:2013nda}). Hence, we can only use the matching of the elliptic genus as a test
of the mass deformed duality.}:
\be
\mathcal{I}_{\mathcal{T}_{\text{A}}}&=&\prod_{i=1}^N\thetafunc{q\,x^{-i}}\oint\frac{\udl{\vec{z}_N}}{\thetafunc{x}^N\prod_{i<j}^N\thetafunc{x\,z_i^{\pm1}z_j^{\pm1}}}\times\nn\\
&\times&\frac{1}{\prod_{i=1}^N\prod_{a=1}^4\thetafunc{s\,x^{\frac{1-N}{3}}u_az_i^{\pm1}}}=\prod_{i=1}^N\frac{\thetafunc{q\,s^{-4}x^{i-\frac{2N+1}{3}}}}{\prod_{a<b}^4\thetafunc{s^2x^{i-\frac{2N+1}{3}}u_au_b}}=\mathcal{I}_{\mathcal{T}_{\text{B}}}\,,
\ee
where we turned on fugacities $u_a$, $s$, $x$ in the Cartan of the global symmetry group $SU(4)_u\times U(1)_s\times U(1)_x$, with $\prod_{a=1}^4u_a=1$. 

We start from the elliptic genus of theory $\mathcal{T}_{\text{A}}$, which from now on we will denote by
\be 
\mathcal{I}_{\mathcal{T}_{\text{A}}}=\mathcal{I}_N({\bf u},s,x)\,,
\ee
and use \eqref{idIP2d} from right to left to replace the contribution of the $USp(2N)$ antisymmetric chiral field $A$ with an auxiliary $(N-1)$-dimensional integral
\be
\mathcal{I}_N({\bf u},s,x)&=&\prod_{i=1}^{N-1}\thetafunc{q\,x^{-i}}\oint\udl{\vec{w}_{N-1}}\oint\frac{\udl{\vec{z}_N}}{\prod_{i=1}^N\prod_{a=1}^4\thetafunc{s\,x^{\frac{1-N}{3}}z_i^{\pm1}}\prod_{\ga=1}^{N-1}\thetafunc{x^{\frac{1}{2}}z_i^{\pm1}w_\ga^{\pm1}}}\,.\nn\\
\ee
This expression can be interpreted as the elliptic genus of the auxiliary dual frame $\mathcal{T}_{\text{aux}}^{(1)}$, where we have a quiver gauge theory with gauge group $USp(2(N-1))\times USp(2N)$. 

Notice that in this auxiliary theory the original $USp(2N)$ node sees $2N+2$ fundamental chirals and no antisymmetric chiral anymore. Hence, we can apply again \eqref{idIP2d} from left to right to confine it
\be
\mathcal{I}_N({\bf u},s,x)&=&\frac{\thetafunc{q\,s^{-4}x^{\frac{N-1}{3}}}}{\prod_{a<b}^4\thetafunc{s^2x^{-\frac{2}{3}(N-1)}u_au_b}}\prod_{i=1}^{N-1}\thetafunc{q\,x^{-i}}\times\nn\\
&\times&\oint\frac{\udl{\vec{w}_{N-1}}}{\thetafunc{x}^{N-1}\prod_{\ga<\gb}^{N-1}\thetafunc{x\,w_\ga^{\pm1}w_\gb^{\pm1}}\prod_{\ga=1}^{N-1}\prod_{a=1}^4\thetafunc{s\,x^{\frac{5-2N}{6}}w_\ga^{\pm1}u_a}}\,.\nn\\
\ee
We now observe that we obtained an integral of the same form of the original one, but of one dimension less, with a different prefactor and with a shift of the parameter $s$. In other words, we obtained the following recursive relation:
\be
\mathcal{I}_N({\bf u},s,x)&=&\frac{\thetafunc{q\,s^{-4}x^{\frac{N-1}{3}}}}{\prod_{a<b}^4\thetafunc{s^2x^{-\frac{2}{3}(N-1)}u_au_b}}\mathcal{I}_{N-1}({\bf u},s\,x^{\frac{1}{6}},x)\,.
\label{recursiverel}
\ee

This relation is extremely powerful. Indeed, as we explained before, what we want to do next is to iterate the two steps we just performed $N$ times, so to completely confine the original $USp(2N)$ integral and get an expression that takes the form of the elliptic genus of a LG model. Equation \eqref{recursiverel} allows us to do so with very little effort. Applying the two previous steps a second time, we get
\be
\mathcal{I}_N({\bf u},s,x)&=&\frac{\thetafunc{q\,s^{-4}x^{\frac{N-1}{3}}}\thetafunc{q\,s^{-4}x^{\frac{N-4}{3}}}}{\prod_{a<b}^4\thetafunc{s^2x^{-\frac{2}{3}(N-1)}u_au_b}\thetafunc{s^2x^{-\frac{1}{3}(2N-5)}u_au_b}}\mathcal{I}_{N-2}({\bf u},s\,x^{\frac{1}{3}},x)=\nn\\
&=&\frac{\prod_{i=N-1}^N\thetafunc{q\,s^{-4}x^{i-\frac{2N+1}{3}}}}{\prod_{i=1}^2\prod_{a<b}^4\thetafunc{s^2x^{i-\frac{2N+1}{3}}u_au_b}}\mathcal{I}_{N-2}({\bf u},s\,x^{\frac{1}{3}},x)\,.
\ee

We can easily get the expression we obtain after $n$ iterations
\be
\mathcal{I}_N({\bf u},s,x)=\frac{\prod_{i=N-n+1}^N\thetafunc{q\,s^{-4}x^{i-\frac{2N+1}{3}}}}{\prod_{i=1}^n\prod_{a<b}^4\thetafunc{s^2x^{i-\frac{2N+1}{3}}u_au_b}}\mathcal{I}_{N-n}({\bf u},s\,x^{\frac{n}{6}},x)\,.
\ee
The complete confinement corresponds to the case $n=N$, for which we find
\be
\mathcal{I}_N({\bf u},s,x)=\prod_{i=1}^N\frac{\thetafunc{q\,s^{-4}x^{i-\frac{2N+1}{3}}}}{\prod_{a<b}^4\thetafunc{s^2x^{i-\frac{2N+1}{3}}u_au_b}}=\mathcal{I}_{\mathcal{T}_{\text{B}}}
\ee
which indeed coincides with the elliptic genus of theory  $\mathcal{T}_{\text{B}}$, as desired.

\section{Duality for $SU(2)$ linear quiver gauge theories}
\label{sec4}

In this section we discuss another duality that can be understood as a generalization of the one for the $SU(2)$ theory with 4  fundamental chirals dual to a LG model. More precisely, we deform the $SU(2)$ theory by introducing two Fermi singlets and propose multiple dual frames for it consisting of $SU(2)$ linear quiver gauge theories of arbitrary length $N-1$ and with $N$ Fermi singlets, where $N$ is any non-negative integer.

We start reviewing the $4d$ ancestor of this duality, then we discuss its dimensional reduction and finally we perform some tests for the resulting $2d$ duality, including a derivation by iterative applications of the most fundamental $N=1$ duality. This derivation is reminiscent of a similar one for the parent $4d$ duality.

\subsection{The $4d$ duality}
\label{secondgen4d}

The four-dimensional duality we want to consider is one of the recently proposed $4d$ mirror-like dualities \cite{Hwang:2020wpd}. More precisely, following the same nomenclature of \cite{Hwang:2020wpd}, we are interested in the duality for the $E^\gs_\gr[USp(2N)]$ theory with the partitions of $N$ $\gr$ and $\gs$ being $\gr=[N-1,1]$ and $\gs=[1^N]$\footnote{For $N=2$
both of the dual theories become $SU(2)$ gauge theories with 8 fundamental chirals and some gauge singlets. This theory is known to have 72 duality frames \cite{Csaki:1997cu} and the duality we are considering here becomes for $N=2$ just a combination of some of these self-dualities. It also corresponds to the self-duality of the $E[USp(4)]$ theory of \cite{Pasquetti:2019hxf} (see also \cite{Garozzo:2020pmz,HPS} for other interesting IR properties of this theory).}. This duality can be considered as a four-dimensional uplift of the $3d$ $\mathcal{N}=4$ abelian mirror duality that relates a $U(1)$ gauge theory with $N$ flavors of fundamental hypermultiplets and a linear abelian quiver with $N-1$ $U(1)$ gauge nodes and one flavor attached to each end of the tail \cite{Intriligator:1996ex}. More precisely, upon dimensional reduction on $\mathbb{S}^1$ and a series of real mass deformations, the $4d$ mirror duality reduces to the $3d$ one. Here we will consider a different dimensional reduction of the $4d$ duality to $2d$.

Let us denote with $\mathcal{T}^{\text{4d}}_{\text{A}}$ and $\mathcal{T}^{\text{4d}}_{\text{B}}$ the two dual four-dimensional theories. The content of theory $\mathcal{T}^{\text{4d}}_{\text{A}}$ can be schematically represented with the following $4d$ $\mathcal{N}=1$ quiver diagram\footnote{As it was already pointed out in \cite{Hwang:2020wpd}, this theory is asymptotically free only for $N<4$, so the $4d$ duality that we are considering becomes a duality between IR free theories if $N$ is too large. Nevertheless, it may still happen that upon compactification it reduces to a duality between interacting theories also for $N>4$. For example, the $3d$ abelian mirror symmetry that we obtain by compactification on $\mathbb{S}^1$ is a duality between interacting theories for any $N$.}

\newpage
\begin{figure}[h]
	\centering
  	\includegraphics[scale=0.6]{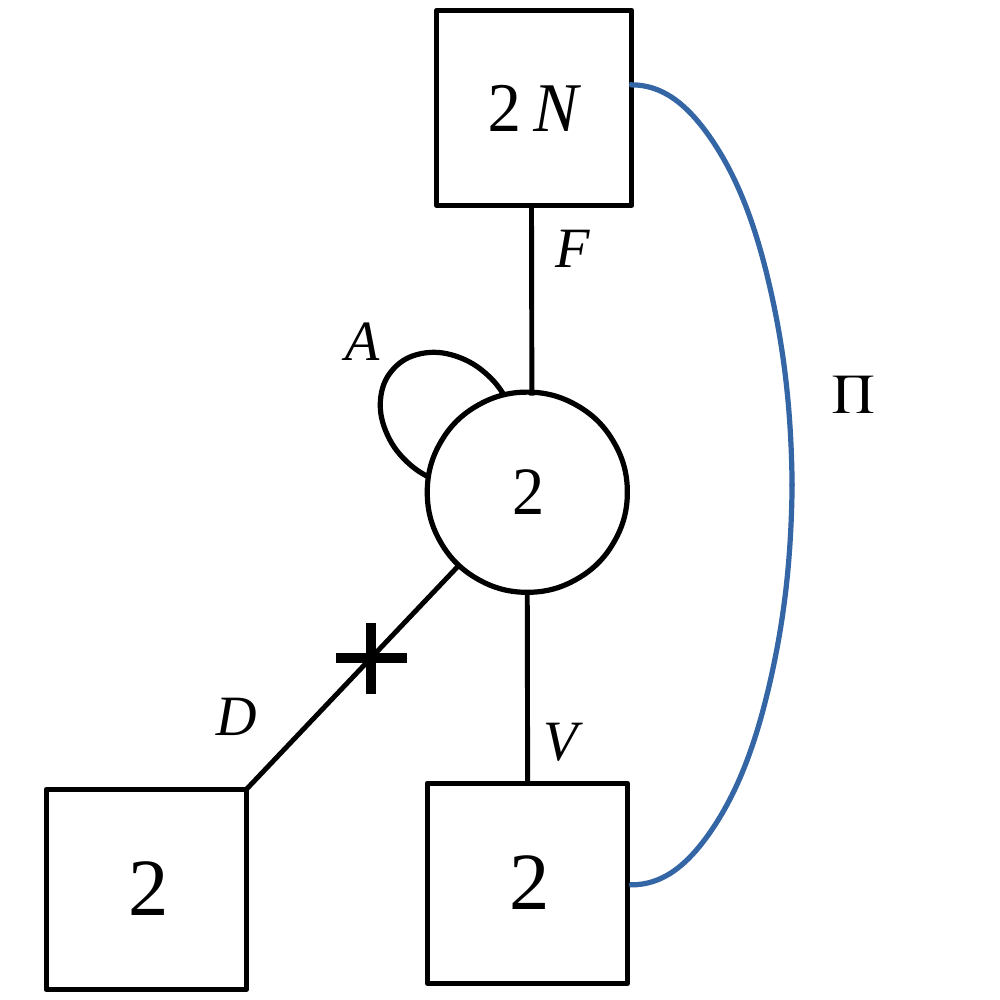} 
\end{figure}

\noindent where all the nodes with a label $2n$ inside correspond to $USp(2n)$ groups, with circular nodes being gauge symmetries and square nodes global symmetries. Moreover, the crosses on some of the straight lines denote gauge singlet fields that are flipping the mesonic operators constructed with the corresponding chirals, while the blue line represents other gauge singlets that are charged under the non-abelian global symmetries. The full superpotential of the theory is
\be
\mathcal{W}_{\mathcal{T}^{\text{4d}}_{\text{A}}}=A\,F^2+F\,\Pi\,V+\ga\,D\,D\,,
\ee
where contractions of gauge and flavor $USp(2n)$ indices are understood.

The dual theory $\mathcal{T}^{\text{4d}}_{\text{B}}$ is a linear quiver of $N-1$ $SU(2)$ gauge groups with the following schematic structure:
\begin{figure}[h]
	\centering
  	\includegraphics[scale=0.6]{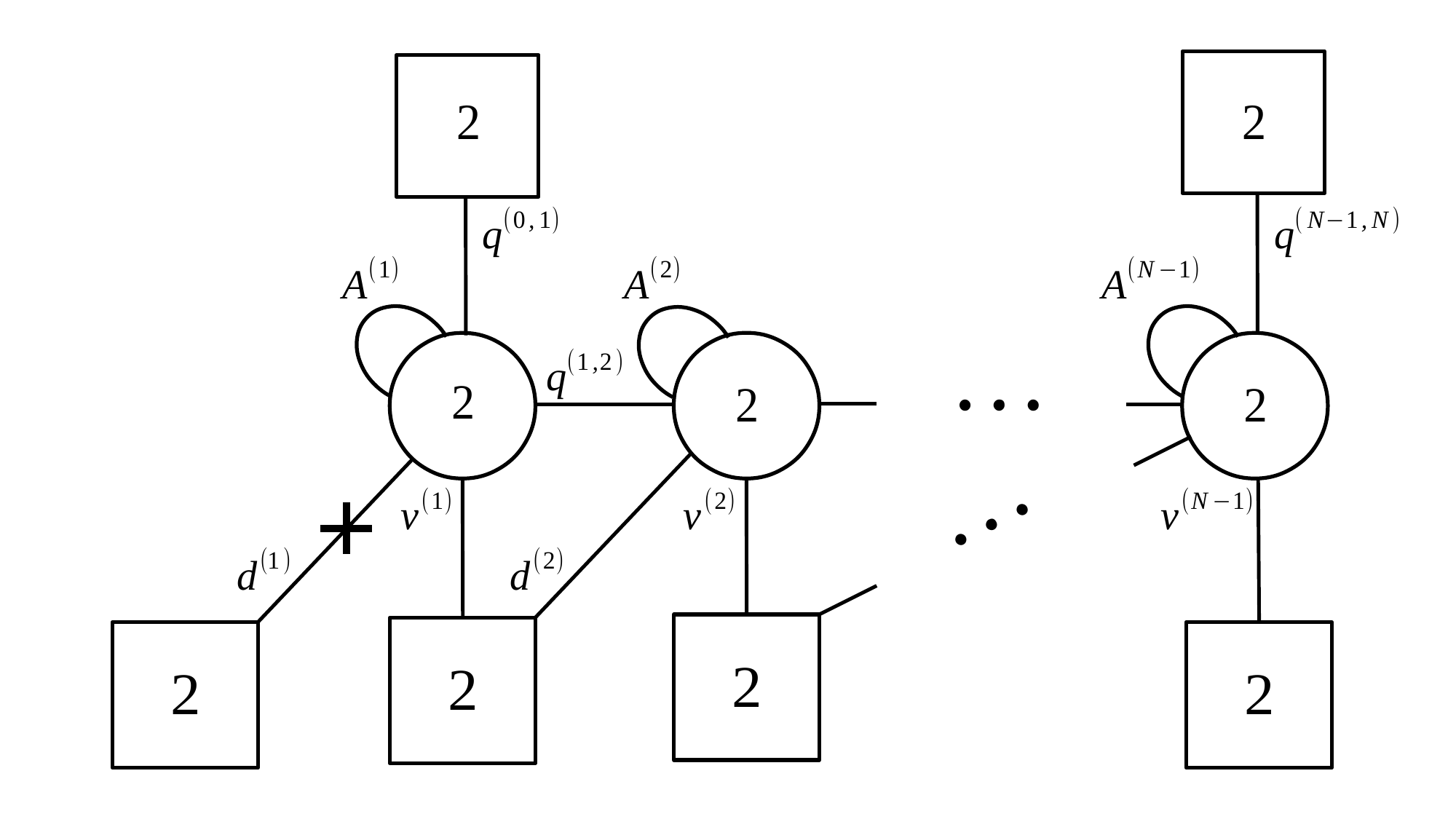} 
\end{figure}

\noindent In the quiver we are not showing some gauge singlet chiral fields charged under the non-abelian global symmetries that are analogues of the field $\Pi$ in theory $\mathcal{T}^{\text{4d}}_{\text{A}}$ to avoid cluttering the drawing. These consists of $N-1$ singlets $\pi^{(i)}$ connecting the upper-left square node with all the lower square nodes except the first one on the left and a singlet $\pi$ connecting the upper-right square node with the lower-right square node. The full superpotential of the theory, including the terms involving these singlets, is
\be
\mathcal{W}_{\mathcal{T}^{\text{4d}}_{\text{B}}}&=&\sum_{i=1}^{N-1}A^{(i)}\left(q^{(i,i+1)}q^{(i,i+1)}-q^{(i-1,i)}q^{(i-1,i)}\right)+\sum_{i=1}^{N-2}v^{(i)}q^{(i,i+1)}d^{(i+1)}+\nn\\
&+&\sum_{i=1}^{N-1}\pi^{(i)}\left(\prod_{j=1}^{i}q^{(j-1,j)}\right)v^{(i)}+\gb\,d^{(1)}d^{(1)}\,,
\label{superpotTB4d}
\ee
where the chiral singlet $\gb$ was again represented with a cross in the previous quiver.

The full IR global symmetry group of the dual theories is
\be
USp(2N)\times SU(2)\times SU(2)\times U(1)_c\times U(1)_t\,,
\label{global4dpre}
\ee
where $U(1)_c$ and $U(1)_t$ are two independent non-anomalous abelian symmetries that can be determined by solving the constraints coming from the superpotential and from the requirement that the NSVZ beta-functions vanish at each gauge node. Notice that the $USp(2N)$ symmetry is enhanced in the IR from the $SU(2)$ symmetries of the saw on the side of theory $\mathcal{T}^{\text{4d}}_{\text{B}}$, while it is completely manifest in the quiver description of theory $\mathcal{T}^{\text{4d}}_{\text{A}}$.

We are actually going to consider the $2d$ reduction of a deformation of this duality. This deformation simply consists of moving the chiral singlet $A$ from theory $\mathcal{T}^{\text{4d}}_{\text{A}}$ to theory $\mathcal{T}^{\text{4d}}_{\text{B}}$ and the chiral singlet $\gb$ from $\mathcal{T}^{\text{4d}}_{\text{B}}$ to theory $\mathcal{T}^{\text{4d}}_{\text{A}}$ by means of a flipping procedure (see footnote 9). These little modifications, in particular the one concerning the field $A$, have an important effect on the global symmetries of the two theories, which now become
\be
SU(2N)\times SU(2)\times SU(2)\times U(1)_c\times U(1)_t\,.
\label{global4d}
\ee
In order to understand why this is the case, we have to analyze more in details the deformations on the two sides of the duality.

On the side of theory $\mathcal{T}^{\text{4d}}_{\text{A}}$, after we remove the field $A$ there is no reason why the chirals $F$ should be rotated by a $USp(2N)$ symmetry rather than an $SU(2N)$ symmetry anymore. This is similar to what happens in the $3d$ version of the theory. Indeed, in the $3d$ $\mathcal{N}=4$ $U(1)$ theory with $N$ fundamental hypers the flavor symmetry is $SU(N)$, but once we remove the adjoint chiral contained in the $3d$ $\mathcal{N}=4$ vector multiplet the two sets of $N$ chirals contained in the hypermultiplets are free to rotate independently and the flavor symmetry is enlarged to $SU(N)^2$. The situation in $4d$ is the same, with the role of the $3d$ $U(1)$ adjoint chiral replaced by the $SU(2)$ antisymmetric chiral $A$. Hence, with this modification the superpotential of the theory becomes
\be
\mathcal{W}_{\mathcal{T}^{\text{4d}}_{\text{A}}}=F\,\Pi\,V+\ga\,D\,D+\gb\,V\,V\,,
\ee
where now the $SU(2N)$ indices are contracted with the standard Kronecker delta. The associated quiver diagram is

\newpage
\begin{figure}[h]
	\centering
  	\includegraphics[scale=0.55]{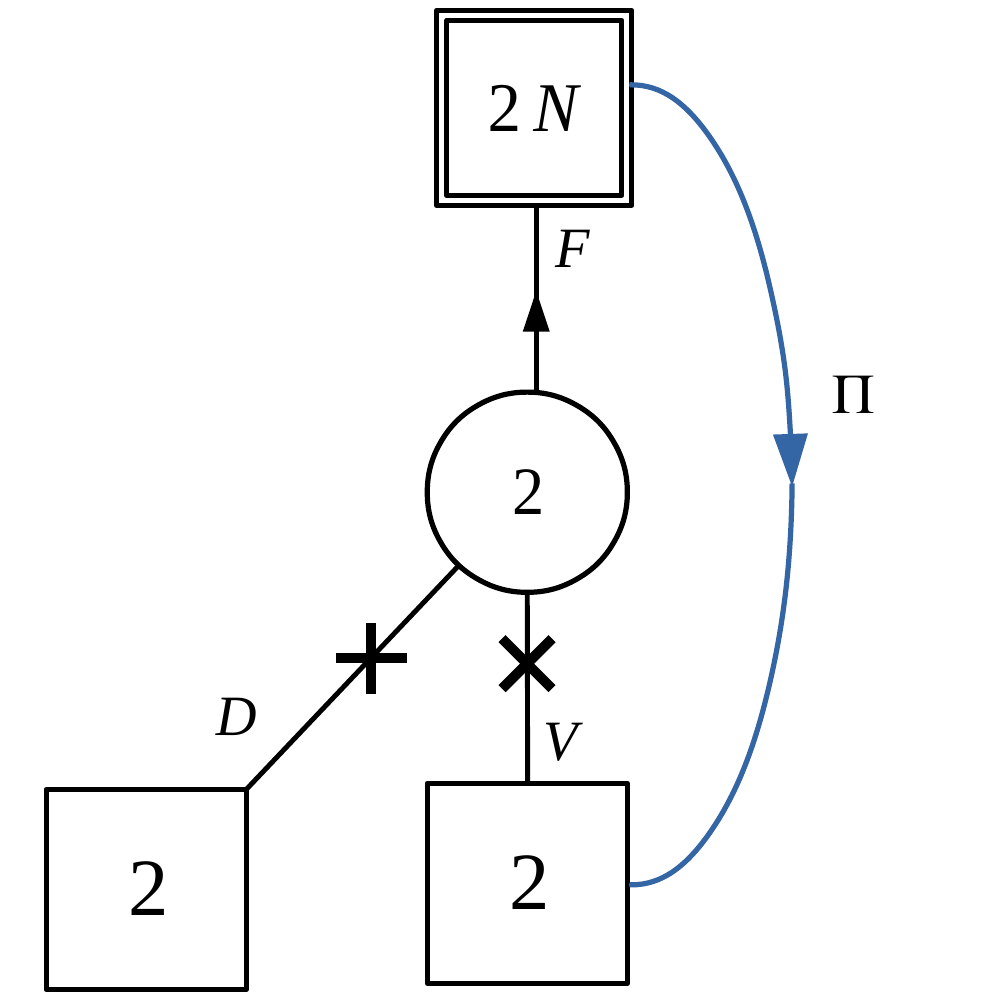} 
\end{figure}

Regarding theory $\mathcal{T}^{\text{4d}}_{\text{B}}$, we first observe that, before the deformation, the chiral $A$ was mapped across the duality to the mesons $q^{(i-1,i)}q^{(i-1,i)}$, which were identified in the chiral ring by the equations of motion of the fields $A^{(i)}$. Hence, we would expect that after the deformation the superpotential of theory $\mathcal{T}^{\text{4d}}_{\text{B}}$ becomes
\be
\mathcal{W}^{\text{unstable}}_{\mathcal{T}^{\text{4d}}_{\text{B}}}&=&\sum_{i=1}^{N-1}A^{(i)}\left(q^{(i,i+1)}q^{(i,i+1)}-q^{(i-1,i)}q^{(i-1,i)}\right)+\sum_{i=1}^{N-2}v^{(i)}q^{(i,i+1)}d^{(i+1)}+\nn\\
&+&\sum_{i=1}^{N-1}\pi^{(i)}\left(\prod_{j=1}^{i}q^{(j-1,j)}\right)v^{(i)}+A\,q^{(0,1)}q^{(0,1)}\,,
\label{superpotTB4dpre}
\ee
This superpotential is actually \emph{unstable} in the sense of the chiral ring stability criterion of \cite{Benvenuti:2017lle}. Consider for example the term $A^{(1)}q^{(0,1)}q^{(0,1)}$. If we remove it from the superpotential, then in the theory that we obtain this operator vanishes in the chiral ring because of the equations of motion of $A$. Similarly, all the terms in the first sum of \eqref{superpotTB4dpre} are unstable, since the F-term equations with respect to the fields $A$, $A^{(i)}$ set all the operators $q^{(i-1,i)}q^{(i-1,i)}$ to zero. 
For this specific case, removing the unstable terms from the superpotential is equivalent to a linear field redefinition of the chiral fields $A$, $A^{(i)}$. Denoting with $B^{(i)}$ for $i=1,\cdots,N$ these new fields, the needed field redefinition is
\be
\begin{cases}
B^{(1)}=A+A^{(1)} \\
B^{(i)}=A^{(i)}-A^{(i-1)} & i=2,\cdots,N-1 \\
B^{(N)}=A^{(N-1)}
\end{cases}
\ee
and the correct stable superpotential reads
\be
\mathcal{W}^{\text{stable}}_{\mathcal{T}^{\text{4d}}_{\text{B}}}&=&\sum_{i=1}^{N-1}B^{(i)}q^{(i-1,i)}q^{(i-1,i)}+\sum_{i=1}^{N-2}v^{(i)}q^{(i,i+1)}d^{(i+1)}+\sum_{i=1}^{N-1}\pi^{(i)}\left(\prod_{j=1}^{i}q^{(j-1,j)}\right)v^{(i)}\,,\nn\\
\label{superpotTB4d}
\ee
The associated quiver diagram is
\begin{figure}[h!]
	\centering
  	\includegraphics[scale=0.55]{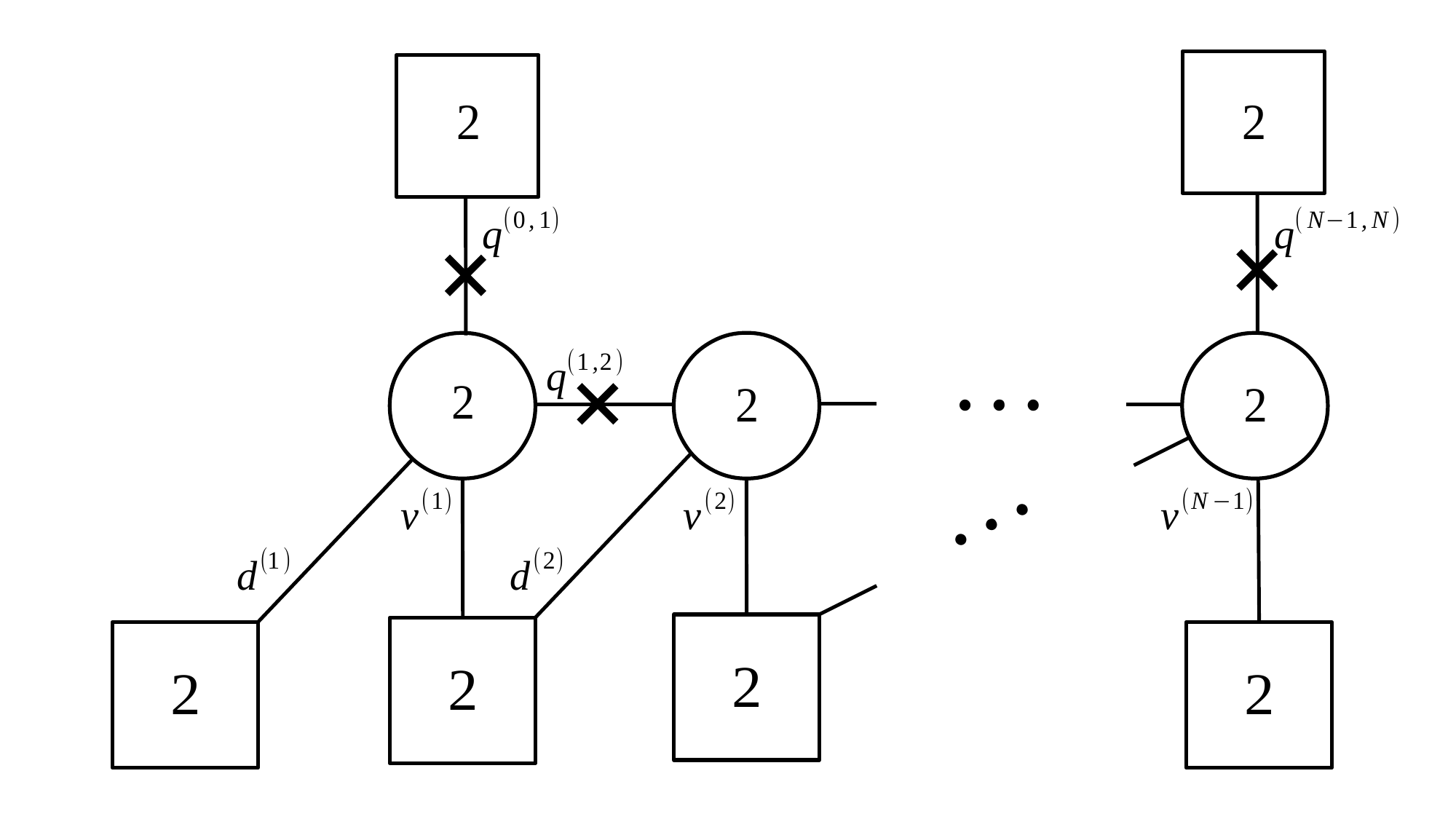} 
\end{figure}

As mentioned before, this is very similar to what happens with the $\mathcal{N}=2$ deformation of the $3d$ $\mathcal{N}=4$ abelian mirror duality where we move the $U(1)$ adjoint chiral on the opposite side of the duality, see Section 2 of \cite{Benvenuti:2016wet}. Also on the side of theory $\mathcal{T}^{\text{4d}}_{\text{B}}$ the global symmetry gets enlarged, as the chirals $B^{(i)}$ are now free to rotate under independent $U(1)$ symmetries. In particular, the manifest global symmetry of the theory is now
\be
\prod_{i=1}^{N}SU(2)_{y_i}\times SU(2)\times SU(2)\times U(1)_c\times U(1)_t\times \prod_{i=1}^{N-1}U(1)_{s_i}\,.
\ee
This symmetry, as it can be understood from the duality\footnote{This new version of the duality can be checked, for example, matching anomalies and the supersymmetric index.}, gets enhanced in the IR to \eqref{global4d}. In particular the symmetries $\prod_{i=1}^{N}SU(2)_{y_i}\times \prod_{i=1}^{N-1}U(1)_{s_i}$ are enhanced to $SU(2N)$.

In order to study the $2d$ reduction of this duality, it will be useful to write explicitly how the chiral fields of the two theories transform under the manifest abelian global symmetries, as this information is not directly represented in the quiver. We are going to specify these data with the following combination:
\be
R=R_0+\sum_{\ga}q_\ga R_\ga\,,
\ee
where $R_0$ represents the R-charge under a trial non-anomalous R-symmetry, $q_\ga$ are the charges under the $\ga$-th abelian symmetry and $R_\ga$ are possible mixing coefficients of the trial R-symmetry with the abelian symmetries. Hence, the R-charge $R$ defined as a function of the mixing coefficients $R_\ga$ encodes all the possible choices of non-anomalous R-symmetries.
Choosing a particular parametrization of the abelian symmetries and of the trial R-symmetry, on the side of theory $\mathcal{T}^{\text{4d}}_{\text{A}}$ we have
\be
\begin{cases}
R[F]=1-\frac{1}{2}R_t\\
R[D]=-1+R_c+\frac{1}{2}R_t \\
R[V]=1-Rc+\frac{N-1}{2}R_t \\
R[\Pi]=R_c -\left(\frac{N}{2}-1\right)R_t \\ 
R[\ga]=4-2R_c-R_t \\
R[\gb]=2R_c-(N-1)R_t
\end{cases}
\label{RchargestheoryA}
\ee
Instead, on the side of theory $\mathcal{T}^{\text{4d}}_{\text{B}}$ we have\footnote{Performing $a$-maximization \cite{Intriligator:2003jj} one finds $R_{s_i}=0$ for every $i$, as expected from the fact that the $U(1)_{s_i}$ symmetries combine with the $SU(2)_{y_i}$ symmetries to enhance to the non-abelian $SU(2N)$ symmetry.}
\be
\begin{cases}
R[q^{(i-1,i)}]=\frac{1}{2}R_t+R_{s_i} & i=1,\cdots,N\\
R[B^{(i)}]=2-R_t-2R_{s_i}& i=1,\cdots,N \\
R[d^{(i)}]=R_c-\frac{N-i}{2}R_t+\sum_{n=1}^{i-1}R_{s_n} & i=1,\cdots,N-1 \\
R[v^{(i)}]=2-R_c +\frac{N-i-2}{2}-\sum_{n=1}^{i+1}R_{s_n} & i=1,\cdots,N-1 \\ 
R[\pi]=R_c+\sum_{n=1}^{N-1}R_{s_n} \\
R[\pi^{(i)}]=R_c-\frac{N-2}{2}R_t+R_{s_{i+1}} & i=1,\cdots,N-1
\end{cases}
\label{RchargestheoryB}
\ee
where we are working in a parametrization of the abelian symmetries in which $\sum_{i=1}^NR_{s_i}=0$.

\subsection{Dimensional reduction}
\label{secondgendimred}

In order to dimensionally reduce this duality, we choose a four-dimensional R-symmetry corresponding to the following values of the 
mixing coefficients $R_c$ and $R_t$:
\be
R_c=1,\qquad R_t=0\,.
\ee
Moreover, we don't mix the R-symmetry with any non-abelian flavor symmetry, meaning that on the side of theory $\mathcal{T}^{\text{4d}}_{\text{B}}$ we also set $R_{s_i}=0$ for all $i$ in \eqref{RchargestheoryB}.
Let us reconstruct the resulting two-dimensional theories separately using the prescription we reviewed in Subsection \ref{dimredgen}. 

For theory $\mathcal{T}^{\text{4d}}_{\text{A}}$ we can see from \eqref{RchargestheoryA} that the fields $F$ and $\Pi$ have R-charge 1 and don't survive the dimensional reduction. The fields $D$ and $V$ have instead R-charge 0 and reduce to two chiral fields in $2d$ which we denote by $L$ and $R$. Finally, the gauge singlets $\ga$ and $\gb$ have R-charge 2 and become Fermi singlets $\Psi_L$ and $\Psi_R$. The resulting theory $\mathcal{T}_{\text{A}}$ can be summarized with the following quiver diagram:
\begin{figure}[h]
	\centering
  	\includegraphics[scale=0.55]{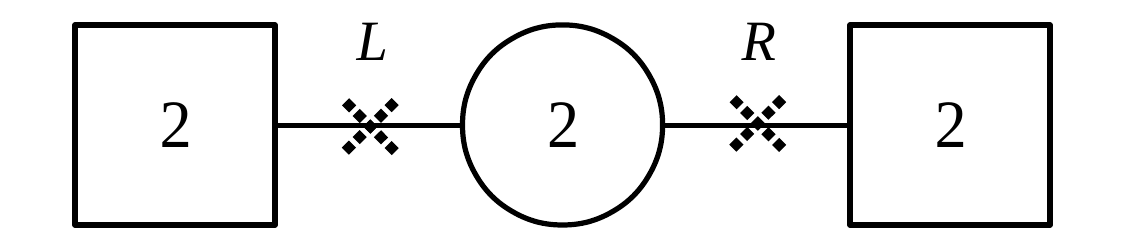} 
\end{figure}

\noindent Of the original $4d$ superpotential only the flipping terms survived
\be
W_{\mathcal{T}_{\text{A}}}=\Psi_LL\,L+\Psi_RR\,R\,.
\ee

\noindent We thus see that all the theories $\mathcal{T}^{\text{4d}}_{\text{A}}$ reduce for any $N$ to the same $2d$ $\mathcal{N}=(0,2)$ theory, namely the $SU(2)$ gauge theory with 4 fundamental chirals and two Fermi singlets. Indeed, since the fields $F$ and $\Pi$ didn't survive the dimensional reduction, the $2d$ theory doesn't possess the $SU(2N)$ factor of the original $4d$ global symmetry \eqref{global4d}, so that the global symmetry of $\mathcal{T}_{\text{A}}$ is $SU(2)_l\times SU(2)_r\times U(1)_c\times U(1)_t$.
The charges of the $2d$ fields under these symmetries inherited from four dimensions are
\begin{table}[h]
\centering
\scalebox{1}{
\setlength{\extrarowheight}{1pt}
\begin{tabular}{c|cccc|c}
{} & $SU(2)_l$ & $SU(2)_r$ & $U(1)_c$ & $U(1)_t$ & $U(1)_{R_0}$ \\ \hline
$L$ & $\Box$ & $\bullet$ & 1 & $\frac{1}{2}$ & $0$ \\
$R$ & $\bullet$ & $\Box$ & $-1$ & $\frac{N-1}{2}$ & $0$ \\
$\Psi_L$ & $\bullet$ & $\bullet$ & $-2$ & $-1$ & $1$ \\ 
$\Psi_R$ & $\bullet$ & $\bullet$ & $2$ & $1-N$ & $1$
\end{tabular}}
\end{table}

\noindent where $U(1)_{R_0}$ denotes the UV trial R-symmetry. It is useful to redefine the abelian symmetries so to completely remove any remnant of the dependence on $N$. At the level of fugacities, the redefinition we want to perform is
\be
d=c\,t^{-\frac{N-2}{4}}\,\qquad s=t^{\frac{N}{4}}\,.
\ee
Equivalently, at the level of the charges we have
\be
\mathcal{Q}_d=\mathcal{Q}_c,\qquad \mathcal{Q}_s=\frac{4\mathcal{Q}_t+(N-2)\mathcal{Q}_c}{N}\,.
\ee
With this choice, the charges of the fields of theory $\mathcal{T}_{\text{A}}$ inherited from four dimensions are
\begin{table}[h]
\centering
\scalebox{1}{
\setlength{\extrarowheight}{1pt}
\begin{tabular}{c|cccc|c}
{} & $SU(2)_l$ & $SU(2)_r$ & $U(1)_d$ & $U(1)_s$ & $U(1)_{R_0}$ \\ \hline
$L$ & $\Box$ & $\bullet$ & 1 & $1$ & $0$ \\
$R$ & $\bullet$ & $\Box$ & $-1$ & $1$ & $0$ \\
$\Psi_L$ & $\bullet$ & $\bullet$ & $-2$ & $-2$ & $1$ \\ 
$\Psi_R$ & $\bullet$ & $\bullet$ & $2$ & $-2$ & $1$
\end{tabular}}
\end{table}

For theory $\mathcal{T}^{\text{4d}}_{\text{B}}$ we can see from \eqref{RchargestheoryB} that all the fields of the saw $d^{(i)}$ and $v^{(i)}$ as well as the singlets $\pi^{(i)}$ and $\pi$ have R-charge 1 and don't survive the dimensional reduction. The bifundamental chirals $q^{(i,i+1)}$ have all R-charge 0 and become chiral multiplets $Q^{(i,i+1)}$, while the singlets $B^{(i)}$ have R-charge 2 and become Fermi gauge singlets $\Psi^{(i)}$. The resulting theory can be summarized with the following quiver diagram:
\begin{figure}[h]
	\centering
  	\includegraphics[scale=0.55]{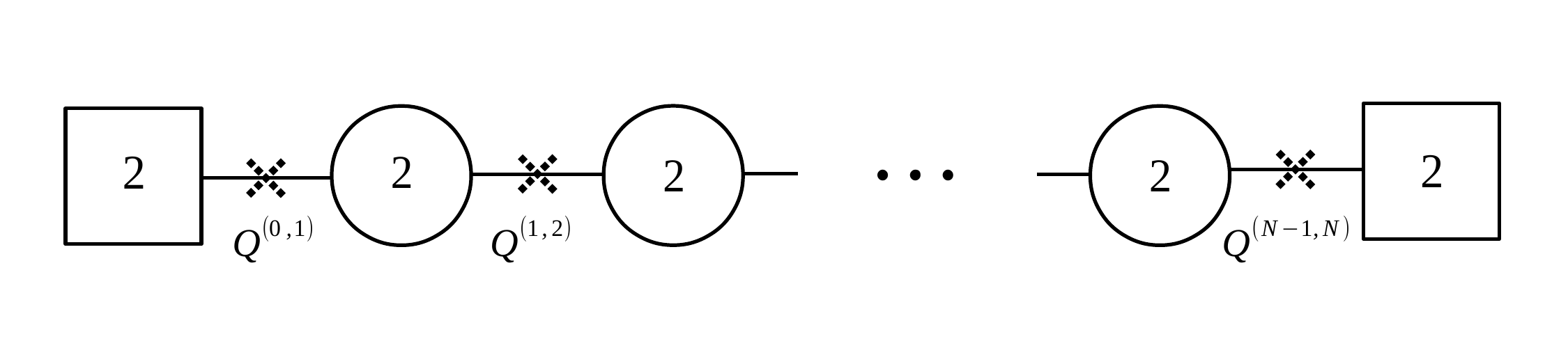} 
\end{figure}

\noindent where the dashed crosses represent the Fermi fields $\Psi^{(i)}$. On this side of the duality we have a non-trivial dependence on $N$ encoded in the length of the quiver and in the number of Fermi singlets. For this reason, we label these theories by $\mathcal{T}_{\text{B}}^{(N)}$.

In this dual frame there is no remnant of the four-dimensional $\prod_{i=1}^NSU(2)_{y_i}$ symmetry. Moreover, all the $4d$ fields charged under $U(1)_c$ didn't survive the dimensional reduction, so that the full manifest global symmetry of the $2d$ theory is only $SU(2)_l\times SU(2)_r\times U(1)_s\times \prod_{i=1}^{N-1}U(1)_{s_i}$.  The charges of all the fields under these symmetries are
\begin{table}[h]
\centering
\scalebox{1}{
\setlength{\extrarowheight}{1pt}
\begin{tabular}{c|cccc|c}
{} & $SU(2)_l$ & $SU(2)_r$ & $U(1)_s$ & $U(1)_{s_i}$ & $U(1)_{R_0}$ \\ \hline
$Q^{(0,1)}$ & $\Box$ & $\bullet$ & $\frac{2}{N}$ & $\gd_{i,1}$ & $0$ \\
$Q^{(j,j+1)}$ & $\bullet$ & $\bullet$ & $\frac{2}{N}$ & $\gd_{i,j+1}$ & $0$ \\
$Q^{(N-1,N)}$ & $\bullet$ & $\Box$ & $\frac{2}{N}$ & $\gd_{i,N}$ & $0$ \\
$\Psi^{(j)}$ & $\bullet$ & $\bullet$ & $-\frac{4}{N}$ & $-2\gd_{i,j}$ & $1$
\end{tabular}}
\end{table}

\noindent where similarly to \eqref{RchargestheoryB} we are using a parametrization of the abelian symmetries such that the combination $\sum_{i=1}^NU(1)_{s_i}=\mathbb{I}$ is the trivial transformation, so in particular $U(1)_{s_N}=-\sum_{i=1}^{N-1}U(1)_{s_i}$.

\subsection{The $2d$ duality}
\label{secondgen2d}

From the dimensional reduction we got the following putative $2d$ $\mathcal{N}=(0,2)$ duality:

\medskip
\noindent \textbf{Theory {\boldmath$\mathcal{T}_{\text{A}}$}}: $SU(2)$ gauge theory with four fundamental chiral fields $L_a$, $R_a$ for $a=1,2$ and two Fermi singlets $\Psi_L$, $\Psi_R$ interacting with
\be
W_{\mathcal{T}_{\text{A}}}=\Psi_LL\,L+\Psi_RR\,R\,.
\ee

\medskip
\noindent \textbf{Theory {\boldmath$\mathcal{T}_{\text{B}^{(N)}}$}}: Linear quiver with $N-1$ $SU(2)$ gauge groups connected by bifundamental chiral fields $Q^{(i,i+1)}$ for $i=1,\cdots,N-2$, with one fundamental chiral at each end of the tail $Q^{(0,1)}$, $Q^{(N-1,N)}$ and $N$ gauge singlet Fermi fields $\Psi^{(i)}$ for $i=1,\cdots,N$ interacting through
\be
W_{\mathcal{T}_{\text{B}}}=\sum_{i=1}^{N-1}\Psi^{(i)}Q^{(i-1,i)}Q^{(i-1,i)}\,.
\ee

Notice that theory $\mathcal{T}_{\text{A}}$ is always the same for any $N$, while theory $\mathcal{T}_{\text{B}}^{(N)}$ changes. This means that from infinitely many dualities in $4d$ we obtained a single duality, but between infinitely many dual frames. We will show that of all of the dualities relating these multiple frames only one is independent, in the sense that all the others can be derived by iterating the duality for $N=1$.

Another peculiarity of this duality is related to the global symmetries of the two theories. We recall that the charges of all the fields that we predict from $4d$ and that we can equivalently determine directly in $2d$ by solving the superpotential constraints are
\begin{table}[h]
\centering
\scalebox{1}{
\setlength{\extrarowheight}{1pt}
\begin{tabular}{c|ccccc|c}
{} & $SU(2)_l$ & $SU(2)_r$ & $U(1)_d$ & $U(1)_s$ & $U(1)_{s_i}$ & $U(1)_{R_0}$ \\ \hline
$L$ & $\Box$ & $\bullet$ & 1 & $1$ & 0 & $0$ \\
$R$ & $\bullet$ & $\Box$ & $-1$ & $1$ & 0 & $0$ \\
$\Psi_L$ & $\bullet$ & $\bullet$ & $-2$ & $-2$ & 0 & $1$ \\ 
$\Psi_R$ & $\bullet$ & $\bullet$ & $2$ & $-2$ & 0 & $1$ \\ \hline
$Q^{(0,1)}$ & $\Box$ & $\bullet$ & 0 & $\frac{2}{N}$ & $\gd_{i,1}$ & $0$ \\
$Q^{(j,j+1)}$ & $\bullet$ & $\bullet$ & 0 & $\frac{2}{N}$ & $\gd_{i,j+1}$ & $0$ \\
$Q^{(N-1,N)}$ & $\bullet$ & $\Box$ & 0 & $\frac{2}{N}$ & $\gd_{i,N}$ & $0$ \\
$\Psi^{(j)}$ & $\bullet$ & $\bullet$ & 0 & $-\frac{4}{N}$ & $-2\gd_{i,j}$ & $1$
\end{tabular}}
\end{table}

\newpage
\noindent where we recall that in our conventions $U(1)_{s_N}=-\sum_{i=1}^{N-1}U(1)_{s_i}$. Notice that $U(1)_d$ is not a symmetry of all the $\mathcal{T}_{\text{B}}^{(N)}$ theories, since no $4d$ field charged under it survived the dimensional reduction.
In the same way, we don't have the symmetries $\prod_{i=1}^{N-1}U(1)_{s_i}$ on the side of theory $\mathcal{T}_{\text{A}}$. We will see momentarily that this is similar to what we discussed at the end of Subsection \ref{review1}.
More precisely, we conjecture that these symmetries decouple in the IR also on the sides of the duality where they are manifest in the UV, in the sense that they are not symmetries of the low energy theory. We will argue this by showing that the dependence on both $U(1)_d$ and $\prod_{i=1}^{N-1}U(1)_{s_i}$ disappear both from the anomalies and from the elliptic genus, which are our main tools for probing the IR dynamics of the theories.

\subsubsection*{The case $N=1$}

For $N=1$ the dual theory $\mathcal{T}_{\text{B}}^{(1)}$ consists only of an $SU(2)_l\times SU(2)_r$ bifundamental chiral field $Q^{(0,1)}$ and a Fermi singlet field $\Psi^{(1)}$, without any gauge group

\vspace{0.5cm}
\hspace{3.14cm}{\Large $\mathcal{T}_{\text{A}}$ \hspace{7.3cm} $\mathcal{T}_{\text{B}}^{(1)}$}
\vspace{-0.3cm}
\begin{figure}[h]
	\centering
  	\includegraphics[scale=0.55]{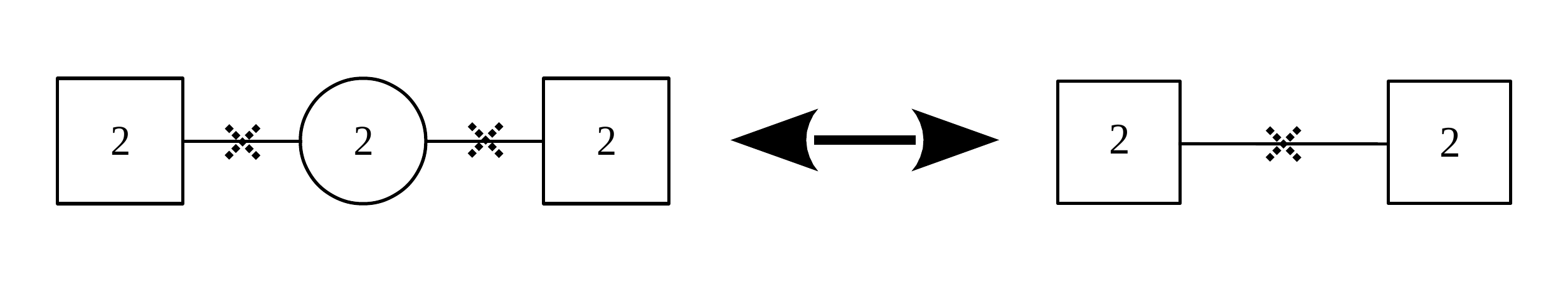} 
\end{figure}

\noindent The interactions on the side of theory $\mathcal{T}_{\text{A}}$ are
\be
W_{\mathcal{T}_{\text{A}}}=\Psi_LL\,L+\Psi_RR\,R\,,
\ee
while on the side of theory $\mathcal{T}_{\text{B}}^{(1)}$
\be
W_{\mathcal{T}_{\text{B}}}=\Psi^{(1)} Q^{(0,1)}Q^{(0,1)}\,.
\label{superpoteTB1}
\ee

\noindent 
This precisely coincides with the alternative version of the duality between $SU(2)$ with 4 chirals and the LG model of 6 chirals and one Fermi fields with a cubic interaction of \cite{Gadde:2015wta,Dedushenko:2017osi} we presented at the end of Subsection \ref{review1}, where we already noticed the decoupling of the $U(1)_d$ symmetry in the IR on the side of theory $\mathcal{T}_{\text{A}}$.

As we mentioned earlier, the duality for generic $N$ can be obtained by applying $N$ times this more fundamental duality.

\subsubsection*{The case $N=2$}

For $N=2$ we don't actually get a duality, as $\mathcal{T}_{\text{A}}$ and $\mathcal{T}_{\text{B}}^{(2)}$ are precisely the same theory. With this, we mean that the gauge group and the matter content are the same, as well as the superpotential interactions. In our notations, we have in this case that the $U(1)_d$ symmetry on the side of theory $\mathcal{T}_{\text{A}}$ coincides with what we called $U(1)_{s_1}$ symmetry on the side of theory $\mathcal{T}_{\text{B}}^{(2)}$. Nevertheless, as we conjectured this symmetry decouples in the IR. The decoupling of the other $U(1)_{s_i}$ symmetries can be understood as an higher $N$ version of the same phenomenon that we have for $U(1)_d$ for $N=1$.

\subsubsection{Anomalies}

A first test of the duality consists of matching the anomalies of $\mathcal{T}_{\text{A}}$ and $\mathcal{T}_{\text{B}}^{(N)}$. On the side of theory $\mathcal{T}_{\text{A}}$ we find that all the abelian anomalies vanish, including the mixed ones. In particular, the vanishing of the anomalies involving $U(1)_d$ is compatible with the decoupling of this symmetry in the IR. 
Similarly, also on the side of theory $\mathcal{T}_{\text{B}}^{(N)}$ all the abelian anomalies vanish, where the vanishing of the $U(1)_{s_i}$ anomalies is again compatible with their decoupling in the IR.

Computing the anomalies for the non-abelian symmetries we find a perfect agreement between theory $\mathcal{T}_{\text{A}}$ and theory $\mathcal{T}_{\text{B}}^{(N)}$, since both of the theories have only two chiral fields transforming in the fundamental representation of each $SU(2)_l$ and $SU(2)_r$ symmetry
\be
&&\Tr\gc^3SU(2)_l^2=2T_{SU(2)_l}(\Box)=1\nn\\
&&\Tr\gc^3SU(2)_r^2=2T_{SU(2)_r}(\Box)=1\,.
\ee

Finally, we compute the trial central charges
\be
c_R=9-24R_s,\qquad c_R-c_L=3\,,
\ee
where $R_s$ is the mixing coefficients of $U(1)_s$ with the trial R-symmetry $U(1)_{R_0}$.  Notice that the result doesn't depend on any mixing coefficient with $U(1)_d$ and $\prod_{i=1}^{N-1}U(1)_{s_i}$, which is again compatible with the decoupling of these symmetries. Moreover, the trial central charge is linear in $R_s$, which doesn't allow us to use $c$-extremization to determine the superconformal R-charge. Once again, this failure in directly applying $c$-extremization is due to the non-compact nature of the target space. Looking at the frame of theory $\mathcal{T}_{\text{B}}^{(1)}$ we can easily see that the classical moduli space is a hypersurface in $\mathbb{C}^4$ defined by the polynomial equation
\be
Q^{(0,1)}Q^{(0,1)}=0\,,
\ee
which comes from the equation of motion of $\Psi^{(1)}$ in \eqref{superpoteTB1}. Hence, we have to require that the chirals $Q^{(0,1)}$ that parametrize the non-compact target space have zero R-charge. This fixes $R_s=0$ and so we get the central charges\footnote{Since in the generic frame $\mathcal{T}_{\text{B}}^{(N)}$ the chirals $Q^{(0,1)}$ of $\mathcal{T}_{\text{B}}^{(1)}$ are mapped to the long mesons $\prod_{i=1}^NQ^{(i-1,i)}$ we also have $R_{s_i}=0$ in these other frames and $R_d=0$ in the frame of $\mathcal{T}_{\text{A}}$. Notice however that this doesn't affect the central charges as they don't depend on these mixing coefficients, in accordance with the fact that these are not symmetries of the low energy theory.}
\be
c_R=9,\qquad c_L=6\,.
\ee

\subsubsection{Elliptic genus and derivation}
\label{EGgen2}

We conclude this section showing how to derive the duality for $N>1$ by iterating the fundamental duality for $N=1$, which is nothing but a deformation of the duality for $SU(2)$ with 4 chirals of \cite{Gadde:2015wta,Dedushenko:2017osi}. 

We remark that the derivation we are going to present is completely analogous to similar derivations for related dualities in higher dimensions. It was first proposed in \cite{Kapustin:1999ha} in the context of $3d$ $\mathcal{N}=4$ abelian mirror symmetry, which is the duality relating a $U(1)$ gauge theory with $N$ fundamental flavors to a linear abelian quiver with $N-1$ nodes and one flavor at each end of the tail \cite{Intriligator:1996ex}. In \cite{Kapustin:1999ha} it was shown how to derive this duality for arbitrary $N$ by applying \emph{piecewise} the duality for $N=1$, which relates the $U(1)$ gauge theory with one fundamental flavor to a free hypermultiplet (see also \cite{Kapustin:2010xq} for an implementation of this procedure at the level of the three-sphere partition function and \cite{Benvenuti:2016wet} for the $\mathcal{N}=2$ case). The four-dimensional duality of which we are considering the $2d$ reduction is an higher dimensional ancestor of this mirror duality and in \cite{Hwang:2020wpd} it was shown that a similar piecewise derivation applies also in $4d$, where the fundamental duality that should be iterated is the Seiberg duality of $SU(2)$ with 6 chirals dual to a WZ model of 15 chirals.

We will describe the piecewise derivation of the $2d$ duality at the level of the elliptic genus. In particular, by applying several times \eqref{idN1alt}, which encodes the duality in the $N=1$ case, we will be able to prove the following integral identity related to the duality for arbitrary $N$:
\begin{equation}
\makebox[\linewidth][c]{\scalebox{0.95}{$
\begin{split}
&\mathcal{I}_{\mathcal{T}_{\text{A}}}=\thetafunc{q\,s^{-2}d^{\pm2}}\oint\frac{\udl{z_1}}{\thetafunc{s\,d\,z^{\pm1}l^{\pm1}}\thetafunc{s\,d^{-1}z^{\pm1}r^{\pm1}}}=\prod_{i=1}^N\thetafunc{q\,s^{-\frac{4}{N}}s_i^{-2}}\times\\
&\times\oint\frac{\prod_{a=1}^{N-1}\udl{z^{(a)}_1}}{\thetafunc{s^{\frac{2}{N}}s_1z^{(1)\pm1}l^{\pm1}}\prod_{a=1}^{N-2}\thetafunc{s^{\frac{2}{N}}s_az^{(a)\pm1}z^{(a+1)\pm1}}\thetafunc{s^{\frac{2}{N}}s_Nz^{(N-1)\pm1}r^{\pm1}}}=\mathcal{I}_{\mathcal{T}_{\text{B}}^{(N)}}\,,
\label{idd}
\end{split}$}}
\end{equation}
where $l$, $r$, $d$, $s$ and $s_i$ are fugacities in the Cartan of the global symmetries $SU(2)_l$, $SU(2)_r$, $U(1)_d$, $U(1)_s$, $\prod_{i=1}^{N-1}U(1)_{s_i}$, with the constraint $\prod_{i=1}^Ns_i=1$. Notice that the elliptic genus of theory $\mathcal{T}_{\text{A}}$ explicitly depends on the fugacity $d$, while that of theory $\mathcal{T}_{\text{B}}^{(N)}$ doesn't. In the same way, the elliptic genus of theory $\mathcal{T}_{\text{B}}^{(N)}$ explicitly depends on the fugacities $s_i$, while that of theory $\mathcal{T}_{\text{A}}$ doesn't. As we pointed out in Subsection \ref{review1}, the identity \eqref{idN1alt}, which corresponds to the case $N=1$ of \eqref{idd}, is valid even without turning off the $d$ fugacity, implying that $\mathcal{I}_{\mathcal{T}_{\text{A}}}$ is actually independent of $d$. Similarly, the identity \eqref{idd} for generic $N$ works even without turning off the fugacities for the $U(1)_d$ and for the $\prod_{i=1}^{N-1}U(1)_{s_i}$ symmetries. The fact that the elliptic genus is independent of the fugacities $d$ and $s_i$ is again compatible with the decoupling of the $U(1)_d$ and $\prod_{i=1}^{N-1}U(1)_{s_i}$ symmetries in the IR, which is necessary in order for the duality to hold.

We start considering the elliptic genus of theory $\mathcal{T}_{\text{B}}^{(N)}$. In particular, we isolate the last $SU(2)$ integral
\be
I^{(1)}=\oint\frac{\udl{z^{(N-1)}_1}}{\thetafunc{s^{\frac{2}{N}}s_{N-1}z^{(N-1)\pm1}z^{(N-2)\pm1}}\thetafunc{s^{\frac{2}{N}}s_Nz^{(N-1)\pm1}r^{\pm1}}}\,.
\ee
Using \eqref{idN1alt} we can rewrite this as
\be
I^{(1)}=\frac{\thetafunc{q\,s^{-\frac{8}{N}}s_{N-1}^{-2}s_N^{-2}}}{\thetafunc{q\,s^{-\frac{4}{N}}s_{N-1}^{-2}}\thetafunc{q\,s^{-\frac{4}{N}}s_{N}^{-2}}\thetafunc{s^{\frac{4}{N}}s_{N-1}s_Nz^{(N-2)\pm1}r^{\pm1}}}\,.
\ee
Plugging this back into the elliptic genus of theory $\mathcal{T}_{\text{B}}^{(N)}$ we get
\be
\mathcal{I}_{\mathcal{T}_{\text{B}}^{(N)}}&=&\thetafunc{q\,s^{-\frac{8}{N}}s_{N-1}^{-2}s_N^{-2}}\prod_{i=1}^{N-2}\thetafunc{q\,s^{-\frac{4}{N}}s_i^{-2}}\times\nn\\
&\times&\oint\frac{\prod_{a=1}^{N-2}\udl{z^{(a)}_1}}{\thetafunc{s^{\frac{2}{N}}s_1z^{(1)\pm1}l^{\pm1}}\prod_{a=1}^{N-3}\thetafunc{s^{\frac{2}{N}}s_az^{(a)\pm1}z^{(a+1)\pm1}}\thetafunc{s^{\frac{4}{N}}s_{N-1}s_Nz^{(N-2)\pm1}r^{\pm1}}}\,.\nn\\
\ee
Now we consider the following integral:
\be
I^{(2)}=\oint\frac{\udl{z^{(N-2)}_1}}{\thetafunc{s^{\frac{2}{N}}s_{N-2}z^{(N-2)\pm1}z^{(N-3)\pm1}}\thetafunc{s^{\frac{4}{N}}s_{N-1}s_Nz^{(N-2)\pm1}r^{\pm1}}}\,.
\ee
Applying the fundamental identity \eqref{idN1alt} again we get
\be
I^{(2)}=\frac{\thetafunc{q\,s^{-\frac{12}{N}}s_{N-2}^{-2}s_{N-1}^{-2}s_N^{-2}}}{\thetafunc{q\,s^{-\frac{4}{N}}s_{N-2}^{-2}}\thetafunc{q\,s^{-\frac{8}{N}}s_{N-1}^{-2}s_N^{-2}}\thetafunc{s^{\frac{6}{N}}s_{N-2}s_{N-1}s_Nz^{(N-3)\pm1}r^{\pm1}}}
\ee
and plugging this into the elliptic genus of $\mathcal{T}_{\text{B}}^{(N)}$ we get
\be
\mathcal{I}_{\mathcal{T}_{\text{B}}^{(N)}}&=&\thetafunc{q\,s^{-\frac{12}{N}}s_{N-2}^{-2}s_{N-1}^{-2}s_N^{-2}}\prod_{i=1}^{N-3}\thetafunc{q\,s^{-\frac{4}{N}}s_i^{-2}}\times\nn\\
&\times&\oint\frac{\prod_{a=1}^{N-3}\udl{z_1^{(a)}}}{\thetafunc{s^{\frac{2}{N}}s_1z^{(1)\pm1}l^{\pm1}}\prod_{a=1}^{N-3}\thetafunc{s^{\frac{2}{N}}s_az^{(a)\pm1}z^{(a+1)\pm1}}\thetafunc{s^{\frac{6}{N}}s_{N-2}s_{N-1}s_Nz^{(N-3)\pm1}r^{\pm1}}}\,.\nn\\
\ee
We want to iterate this procedure $N-1$ times. At the $n$-th iteration we have to use the following evaluation formula, which again follows from \eqref{idN1alt}:
\be
I^{(n)}&=&\oint\frac{\udl{z^{(N-n)}_1}}{\thetafunc{s^{\frac{2}{N}}s_{N-n}z^{(N-n)\pm1}z^{(N-n-1)\pm1}}\thetafunc{s^{\frac{2n}{N}}\prod_{i=N-n+1}^Ns_iz^{(N-n)\pm1}r^{\pm1}}}=\nn\\
&=&\frac{\thetafunc{q\,s^{-\frac{4(n+1)}{N}}\prod_{i=N-n}^Ns_i^{-2}}}{\thetafunc{q\,s^{-\frac{4}{N}}s_{N-n}}\thetafunc{q\,s^{-\frac{4n}{N}}\prod_{i=N-n+1}^Ns_i^{-2}}\thetafunc{s^{\frac{2(n+1)}{N}}\prod_{i=N-n}^Ns_iz^{(N-n-1)\pm1}r^{\pm1}}}\,,\nn\\
\ee
where for $n=N-1$ we have $z^{(0)}=l$. Hence, after the $(N-1)$-th iteration we get
\be
\mathcal{I}_{\mathcal{T}_{\text{B}}^{(N)}}=\frac{\thetafunc{q\,s^{-4}\prod_{i=1}^Ns_i^{-2}}}{\thetafunc{s^2\prod_{i=1}^Ns_il^{\pm1}r^{\pm1}}}=\mathcal{I}_{\mathcal{T}_{\text{B}}^{(1)}}\,,
\ee
which coincides with the elliptic genus of theory $\mathcal{T}_{\text{B}}^{(1)}$ since $\prod_{i=1}^Ns_i=1$. Using \eqref{idN1alt} one last time but from right to left, we then get precisely \eqref{idd}.

\section{Conclusions}

In this paper we studied the dimensional reduction of some $4d$ $\mathcal{N}=1$ IR dualities on $\mathbb{S}^2$ with suitable topological twists and argued that the resulting $2d$ $\mathcal{N}=(0,2)$ theories are still dual to each other, provided that massive deformations that lift any possible non-compact direction in the target space are turned on. 

There are still many open questions about this kind of dimensional reduction of dualities, for which it would be interesting to find an answer. In particular, it is not clear why certain dualities survive the dimensional reduction while others don't. From the examples of \cite{Gadde:2015wta} and those studied in the present paper, it seems that everytime one of the two $4d$ dual theories is a WZ model the duality survives the dimensional reduction to $2d$. 

Instead, again thinking of the examples studied so far, it seems that many of the $4d$ self-dualities, namely dualities between theories with same gauge group and same gauge charged matter, but possibly a different number of gauge singlets interacting with different superpotentials, reduce in $2d$ to trivial dualities, in the sense that on both sides of the duality we get the same theory, including the same gauge singlets. This happens in the self-dual case of Intriligator--Pouliot duality \cite{Gadde:2015wta} and it turns out to be true, for example, also for the  dimensional reduction of the 72 dual frames of the $USp(2N)$ gauge theory with one antisymmetric and eight fundamental chirals \cite{Csaki:1997cu,Spiridonov:2008zr,Dimofte:2012pd,Razamat:2017hda,HPS}, which differ for gauge singlets that flip all the possible combinations of mesons and baryons. After the dimensional reduction, all the gauge singlets become massive and the 72 different dual frames collapse to the same one in $2d$.

Nevertheless, it may happen that starting from a different configuration of singlets in the $4d$ self-duality, one can obtain a non-trivial duality in $2d$. This happens for example in the duality of Section \ref{sec4}. For $N=2$ the original $4d$ mirror-like duality of \cite{Hwang:2020wpd} coincides with a combination of some of the 72 dualities enjoyed by $SU(2)$ with 8 chirals, but with a different disposition of singlets which is essential in order for the two theories to enjoy a global symmetry enhancement that makes their symmetries match. This is actually a quite general phenomenon. When we have a self-duality, we can try to find an equivalent version of it where we also have the same number of singlets on both sides of the duality. In such cases, the theory remains invariant under the action of the duality, but this acts non-trivially on the gauge invariant operators of the theory, implying that the Weyl group of the global symmetry of the theory is larger than the one manifest from the Lagrangian. This may lead to an enhancement of the global symmetry in the IR (see \cite{Razamat:2017hda,Razamat:2017wsk,Benini:2018bhk,Razamat:2018gbu,HPS} for some examples). It might be that different compactifications of a $4d$ theory with IR symmetry enhancement that are related by some Weyl element of the enhanced symmetry, which we can understand as a self-duality, reduce to non-trivial $2d$ dualities. It would also be interesting to investigate if these enhancements survive, at least partially, the dimensional reduction \cite{PRS}.

Another interesting possible line of future research is related to the strict analogy that there is between some of the $2d$ $\mathcal{N}=(0,2)$ dualities discussed in this paper and similar dualities between $3d$ $\mathcal{N}=2$ theories. For example, the duality discussed in Section \ref{sec3} has a direct three-dimensional analogue, which was obtained in \cite{Benvenuti:2018bav,Amariti:2018wht} as an $\mathbb{S}^1$ compactification of the same four-dimensional confining duality we started from \cite{Csaki:1996eu}, followed by a suitable real mass deformation. It would be interesting to understand if our duality can be derived also as a dual boundary condition of the duality of \cite{Benvenuti:2018bav,Amariti:2018wht} in the same spirit of \cite{Dimofte:2017tpi} and, more in general, if the dimensional reduction of $4d$ $\mathcal{N}=1$ dualities on $\mathbb{S}^2$ with a topological twist can be reinterpreted as a compactification on $\mathbb{S}^1$ giving a $3d$ $\mathcal{N}=2$ duality, followed by real mass deformations and by the introduction of a boundary condition preserving $2d$ $\mathcal{N}=(0,2)$ supersymmetry.

\section*{Acknowledgements}
We would like to thank Antonio Amariti, Sergio Benvenuti, Ivan Garozzo, Noppadol Mekareeya, Sara Pasquetti and Shlomo Razamat for helpful comments and discussions. We especially thank Ivan Garozzo, Noppadol Mekareeya and Sara Pasquetti for useful suggestions on the draft. M.S. is partially supported by the ERC-STG grant 637844-HBQFTNCER, by the University of Milano-Bicocca grant 2016-ATESP0586, by the MIUR-PRIN contract 2017CC72MK003 and by the INFN. 

\appendix

\section{Another duality for $USp(2N)$ gauge theory with antisymmetric}
\label{appA}

In this appendix we comment on the possibility of another duality for a $USp(2N)$ gauge theory with one antisymmetric chiral, but a different number of fundamental chirals and Fermis with respect to the one discussed in Section \ref{sec3}, more precisely $N_b=5$ and $N_f=1$.

We start again from the $4d$ duality of \cite{Csaki:1996eu} relating a $USp(2N)$ gauge theory with one antisymmetric, 6 fundamental chirals and $N$ chiral singlets to a WZ model of $15N$ chirals with cubic superpotential, but we choose the R-symmetry $U(1)_R^{\text{4d}}$ in a different way than what we have done in Subsection \ref{firstgendimred}. Specifically, we look at the subgroup $SU(3)\times SU(2)\times U(1)_B\times U(1)_p\times U(1)_x\subset SU(6)_v\times U(1)_x$ of the non-anomalous global symmetry and define $U(1)_R^{\text{4d}}$ allowing for a mixing with $U(1)_B$, $U(1)_p$ and $U(1)_x$
\be
R=R_0+q_BR_B+q_pR_p+q_xR_x\,.
\ee
In order to determine the $U(1)_R^{\text{4d}}$ charge of all the chiral fields of the two dual theories, we need to use the following branching rules for the fundamental and antisymmetric representations of $SU(6)_v$ under the $SU(3)\times SU(2)\times U(1)_B\times U(1)_p$ subgroup
\be
&&{\bf 6}\to({\bf 3},{\bf 1})^{(-1,0)}\oplus({\bf 1},{\bf 2})^{(1,1)}\oplus({\bf 1},{\bf 1})^{(1,-2)}\nn\\
&&{\bf 15}\to({\bf 3},{\bf 2})^{(0,1)}\oplus({\bf 3},{\bf 1})^{(0,-2)}\oplus({\bf 1},{\bf \bar{3}})^{(-2,0)}\oplus({\bf 2},{\bf 1})^{(2,-1)}\oplus({\bf 1},{\bf 1})^{(2,2)}\,.
\ee
We can see that we can make the 6 $4d$ chirals become 5 $2d$ chirals and one Fermi by choosing the R-symmetry $U(1)_R^{\text{4d}}$ corresponding to the values of the mixing coefficients $R_p=-\frac{2}{3}$ and $R_B=\frac{1}{3}$. We also choose $R_x=0$ in order to make the $4d$ antisymmetric chiral field $A$ a $2d$ antisymmetric chiral.

With this choice, the $4d$ gauge theory $\mathcal{T}_A^{\text{4d}}$ became a $2d$ $USp(2N)$ gauge theory with one antisymmetric chiral field, $3+2$ fundamental chiral fields, one fundamental Fermi field and $N$ Fermi singlets. The $4d$ WZ theory $\mathcal{T}_B^{\text{4d}}$ instead reduced to a LG model of $(6+3+1)N$ chirals and $(3+2)N$ Fermis. 

It is useful to redefine the abelian symmetries $U(1)_B$ and $U(1)_s$ as
\be
U(1)_d=-2U(1)_B-U(1)_p,\qquad U(1)_s=-U(1)_B+2U(1)_s\,.
\ee
Then the transformation rules of the $2d$ matter fields of the two theories under the global symmetry that we expect from $4d$ are
\begin{table}[h]
\centering
\scalebox{1}{\setlength{\extrarowheight}{2pt}
\begin{tabular}{c|ccccc}
{} & $SU(3)$ & $SU(2)$ & $U(1)_d$ & $U(1)_s$ & $U(1)_x$ \\ \hline
$Q$ & $\bf 3$ & $\bullet$ & 2 & 1 & $\frac{1-N}{3}$ \\
$P$ & $\bullet$ & $\bf 2$ & $-3$ & 1 & $\frac{1-N}{3}$ \\
$\Psi$ & $\bullet$ & $\bullet$ & $0$ & $-5$ & $\frac{1-N}{3}$ \\
$A$ & $\bullet$ & $\bullet$ & 0 & 0 & 1 \\
$\gb_i$ & $\bullet$ & $\bullet$ & 0 & 0 & $-i$ \\ \hline
$\Gp^{(1)}_i$ & $\bf 3$ & $\bf 2$ & $-1$ & $2$ & $i-\frac{2N+1}{3}$ \\
$\Gp^{(2)}_i$ & $\bf \bar{3}$ & $\bullet$ & $4$ & $2$ & $i-\frac{2N+1}{3}$ \\
$\Gp^{(3)}_i$ & $\bullet$ & $\bullet$ & $-6$ & $2$ & $i-\frac{2N+1}{3}$ \\
$\Psi^{(1)}_i$ & $\bf 3$ & $\bullet$ & $2$ & $-4$ & $i-\frac{2N+1}{3}$ \\
$\Psi^{(2)}_i$ & $\bullet$ & $\bf 2$ & $-3$ & $-4$ & $i-\frac{2N+1}{3}$ \\
\end{tabular}}
\end{table}

\noindent
We can thus see that the true global symmetry that is manifest at the Lagrangian level is actually $SU(5)_u\times U(1)_s\times U(1)_x$, where $SU(5)_u$ is enhanced from the $4d$ symmetry $SU(3)\times SU(2)\times U(1)_d$ according to the branching rules
\be
&&{\bf 5}\to({\bf 3},{\bf 1})^{2}\oplus({\bf 1},{\bf 2})^{-3}\nn\\
&&{\bf 10}\to({\bf 3},{\bf 2})^{-1}\oplus({\bf \bar{3}},{\bf 1})^{4}\oplus({\bf 1},{\bf 1})^{-6}\,.
\ee

Summarizing, we get the following putative $2d$ $\mathcal{N}=(0,2)$ duality:

\medskip
\noindent \textbf{Theory \boldmath$\mathcal{T}_{\text{A}}$}: $USp(2N)$ gauge theory with one antisymmetric chiral $A$, five fundamental chirals $Q_a$, one funamental Fermi $\Psi$ and $N$ Fermi singlets $\beta_i$ with superpotential
\be
\mathcal{W}_{\mathcal{T}_{\text{A}}}=\sum_{i=1}^N\gb_i \Tr_NA^i\, .
\ee

\medskip
\noindent \textbf{Theory \boldmath$\mathcal{T}_{\text{B}}$}: LG model with $5N$ Fermi fields $\Psi_{a;i}$ and $10N$ chiral fields $\Gp_{ab;i}$ for $i=1,\cdots,N$, $a<b=1,\cdots,5$ with cubic superpotential
\be
\mathcal{W}_{\mathcal{T}_{\text{B}}}=\sum_{i,j,k=1}^N\sum_{a,b,c,d,e=1}^5\epsilon_{abcde}\Psi_{a;i}\Gp_{bc;j}\Gp_{de;k}\gd_{i+j+k,2N+1}\, .
\ee

Notice that for $N=1$ this duality corresponds to the dimensional reduction of Seiberg duality discussed in eq.~(3.4) of \cite{Gadde:2015wta} in the particular case $N_c=2$, $N_f=3$ and $n=0$.

The global symmetry for the two theories and the transformation rules of the matter fields that we expect from four dimensions are
\begin{table}[h]
\centering
\scalebox{1}{\setlength{\extrarowheight}{2pt}
\begin{tabular}{c|ccc|c}
{} & $SU(5)_u$ & $U(1)_s$ & $U(1)_x$ & $U(1)_{R_0}$ \\ \hline
$Q$ & $\bf 5$ & 1 & $\frac{1-N}{3}$ & 0 \\
$\Psi$ & $\bullet$ & $-5$ & $\frac{1-N}{3}$ & 1 \\
$A$ & $\bullet$ & 0 & 1 & 0 \\
$\gb_i$ & $\bullet$ & 0 & $-i$ & 1 \\ \hline
$\Psi_i$ & $\bf 5$ & $-4$ & $i-\frac{2N+1}{3}$ & 1 \\
$\Gp_i$ & $\bf 10$ & $2$ & $i-\frac{2N+1}{3}$ & 0 \\
\end{tabular}}
\end{table}

\noindent Notice that, from a purely $2d$ point of view, on the side of theory $\mathcal{T}_{\text{A}}$ we would expect an additional symmetry that instead we don't get from $4d$. Indeed, there is apparently no superpotential term that prevents the fundamental chirals $Q$ and the fundamental Fermi $\Psi$ to rotate under two independent $U(1)$ symmetries, while from $4d$ we only get one particular combination of these two symmetries, which we called $U(1)_s$. This is one of the problems that may affect the dimensional reduction of $4d$ $\mathcal{N}=1$ dualities to $2d$ $\mathcal{N}=(0,2)$ that we mentioned in the introduction\footnote{Indeed, as we already pointed out before, our proposed duality reduces for $N=1$ to one of those discussed in \cite{Gadde:2015wta}, which also suffers of the same issue.}. It is essentially due to the different nature of anomalies in $4d$ and $2d$, which makes it possible that a $U(1)$ symmetry that was anomalous in four dimensions is not anomalous anymore in two dimensions. As also mentioned in \cite{Gadde:2015wta}, the $2d$ duality typically holds only if this symmetry is broken also in $2d$. This situation is reminiscent of the perturbatively generated monopole superpotential in the reduction from $4d$ to $3d$, which explicitly breaks $U(1)$ symmetries that were anomalous in $4d$ but which wouldn't be in $3d$. In \cite{Gadde:2015wta} it was suggested that a similar mechanism may come into play also in the $4d$ to $2d$ reduction, but at the moment it is not understood what the analogue of monopole operators should be. It would be interesting to understand this fact more in details.

One test that we can perform for the validity of the duality is matching anomalies. For both of the theories we find the following abelian anomalies:
\begin{equation}
\makebox[\linewidth][c]{\scalebox{0.95}{$
\begin{split}
\Tr\,\gc^3U(1)_s^2=-40N,\quad \Tr\,\gc^3U(1)_x^2=\frac{5}{18} N(N-1) (2 N+1),\quad \Tr\,\gc^3U(1)_sU(1)_x=-\frac{20}{3}N(N-1)
\end{split}$}}
\end{equation}
and the following non-abelian anomaly
\be
\Tr\,\gc^3SU(5)_u^2=N\,.
\ee

We can also match the trial central charges
\be
&&c_R=\frac{5}{6} N \left((N-1) R_x ((2 N+1) R_x+18)-48 (N-1) R_s R_x-144 R_s^2+18\right)\nn\\
&&c_R-c_L=5N\,.
\ee
If we naively perform $c$-extremization, we get
\be
R_s^{\text{naive}}=\frac{N-1}{2(2N-1)},\qquad R_x^{\text{naive}}=-\frac{3}{2N-1}\,,
\ee
which leads to the following values for the superconformal central charges:
\be
c_R^{\text{naive}}=\frac{15N(N+1)}{2(2N-1)}, \qquad c_L^{\text{naive}}=-\frac{5N(N-5)}{2(2N-1)}\,.
\ee
Similarly to what happened for the duality of Subsubsection \ref{anomaliessec3}, the central charges are not positive for any $N$. This is due to the fact that the target space, parametrized by the chirals $\Gp_{ab;i}$, is non-compact. Indeed, on the side of theory $\mathcal{T}_{\text{B}}$ the equations of motion of the Fermi fields $\Psi_{a;i}$ imply that the classical moduli space is a co-dimension $5N$ subspace of $\mathbb{C}^{10N}$ defined by the following $N$ independent polynomial equations:
\be
\sum_{j=0}^{N-1}\sum_{b,c,d,e=1}^5\epsilon_{abcde}\Gp_{bc;N-j}\Gp_{de;N-i+j+1}=0,\qquad i=1,\cdots,N,\quad a=1,\cdots,5\,,
\ee
where we defined $\Gp_i=0$ for all $i>N$. Hence, in order to correctly determine the central charges we have to require that these chiral fields $\Gp_{ab;i}$ have vanishing R-charges. This uniquely fixes the two mixing coefficients to the values
\be
R_s=R_x=0
\ee
and, consequently, the central charges to
\be
c_R=15N,\qquad c_L=10N\,,
\ee
which are now positive for any $N$.

For the potential duality we are discussing in this appendix, differently from the one presented in Section \ref{sec3}, we are not able to provide a derivation using some more fundamental duality. Consequently, we are not able to analytically prove the equality of the elliptic genera of the two theories, which is
\be
\mathcal{I}_{\mathcal{T}_{\text{A}}}&=&\prod_{i=1}^N\thetafunc{q\,x^{-i}}\oint\frac{\udl{\vec{z}_N}}{\thetafunc{x}^N\prod_{i<j}^N\thetafunc{x\,z_i^{\pm1}z_j^{\pm1}}}\times\nn\\
&\times&\prod_{i=1}^N\frac{\thetafunc{q\,s^{-5}x^{\frac{1-N}{3}}z_i^{\pm1}}}{\prod_{a=1}^N\thetafunc{s\,x^{\frac{1-N}{3}}u_az_i^{\pm1}}}=\prod_{i=1}^N\frac{\prod_{a=1}^5\thetafunc{q\,s^{-4}x^{i-\frac{2N+1}{3}}}u_a}{\prod_{a<b}^5\thetafunc{s^2x^{i-\frac{2N+1}{3}}u_au_b}}=\mathcal{I}_{\mathcal{T}_{\text{B}}}\,.
\ee
Unfortunately, also a numerical test of this identity is extremely hard from a computational point of view for any $N$ which is not $N=1$. We only managed to test it for $N=2$ to low orders in a double expansion in either $q$, $s$ or $q$, $x$.

\section{Elliptic genus conventions}
\label{appB}

In this appendix we explain our conventions for the elliptic genus of $2d$ $\mathcal{N}=(0,2)$ theories. The elliptic genus was originally studied in \cite{Schellekens:1986yj,Schellekens:1986yi,Pilch:1986en,Witten:1986bf,Witten:1987cg}. More recently, it has been computed for generic $2d$ $\mathcal{N}=(0,2)$ supersymmetric gauge theories in the NSNS sector in \cite{Gadde:2013wq} (see also \cite{Gadde:2013ftv} for the case of $\mathcal{N}=(2,2)$ supersymmetry) and in the RR sector in \cite{Benini:2013nda,Benini:2013xpa}, where in the last two references it was computed as a partition function on $\mathbb{T}^2$ with localization techniques. We will follow the conventions of \cite{Gadde:2013wq,Gadde:2013ftv} and define it in radial quantization as
\be
\mathcal{I}({\bf u};q)=\Tr_{\text{NSNS}}(-1)^Fq^{L_0}\prod_{a}u_a^{f_a}\,.
\ee
This can be understood as a refined version of the Witten index, where we also turned on fugacities $u_a$ in the Cartan of the global symmetry group $F$, whose corresponding generators we denoted by $f_a$. The parameter $q$ can also be interpreted as $q=\e^{2\pi i\tau}$, where $\tau$ is the complex structue of the torus. 

The elliptic genus has the remarkable property of being independent of the coupling constants of the theory. This allows us to compute it in the free field limit. By doing so, we can equivalently write it in the following integral form:
\be
\mathcal{I}({\bf u};q)=\frac{1}{|W|}\oint\prod_{i=1}^{\mathrm{rk}\,G}\frac{\udl{z_i}}{2\pi iz_i}\mathcal{I}_{\text{vec}}({\bf z};q)\mathcal{I}_{\text{chir}}({\bf z};{\bf u};q)\mathcal{I}_{\text{ferm}}({\bf z};{\bf u};q)\,,
\label{EGgeneral}
\ee
where $G$ denotes the gauge group, $|W|$ is the dimension of its Weyl group, $\mathrm{rk}\,G$ is its rank and $z_i$ are fugacities taking values in its Cartan subalgebra. 

The integrand receives contributions from all the possible multiplets of the theory. Specifically, a chiral multiplet of R-charge $R$ in the representation $\mathcal{R}_G$ of the gauge symmetry group $G$ with weight vectors $\gr$ and $\mathcal{R}_F$ of the global symmetry group $F$ with weight vectors $\tilde{\gr}$ contributes as
\be
\mathcal{I}_{\text{chir}}({\bf z};{\bf u};q)=\prod_{\gr\in\mathcal{R}_G}\prod_{\tilde{\gr}\in\mathcal{R}_F}\frac{1}{\thetafunc{q^{\frac{R}{2}}{\bf z}^\gr {\bf u}^{\tilde{\gr}}}}\,,
\ee
where $\thetafunc{x}=\qfac{x}\qfac{q\,x^{-1}}$, $\qfac{x}=\prod_{k=0}^\infty(1-x\, q^k)$ and we also introduced the short-hand notation ${\bf z}^\gr=\prod_{i=1}^{\mathrm{rk}\,G}z_i^{\gr_i}$. Instead, a Fermi multiplet contributes as
\be
\mathcal{I}_{\text{ferm}}({\bf z};{\bf u};q)=\prod_{\gr\in\mathcal{R}_G}\prod_{\tilde{\gr}\in\mathcal{R}_F}\thetafunc{q^{\frac{R+1}{2}}{\bf z}^\gr {\bf u}^{\tilde{\gr}}}\,.
\ee
Finally, a vector multiplet contributes as
\be
\mathcal{I}_{\text{vec}}({\bf z};q)=\qfac{q}^{2\mathrm{rk}\,G}\prod_{\ga\in\mathfrak{g}}\thetafunc{{\bf z}^\ga}\,,
\ee
where $\ga$ are the roots of the gauge algebra $\mathfrak{g}$.

The integrand has the important property of being an elliptic function, \emph{i.e.} invariant under rescaling $z_i\to q\,z_i$, provided that all the gauge anomalies of the theory vanish. This allows us compute the integral \eqref{EGgeneral} considering poles in the fundamental domain only and neglecting all the multiple copies of poles of the theta-functions. The integration contour is defined according to the Jeffrey--Kirwan residue prescription \cite{1993alg.geom..7001J} (see \cite{Benini:2013xpa} for a detailed explanation). In the case of a gauge theory with fundamental matter only, this amounts to considering all possible $\mathrm{rk}\,G$ simultaneous poles, one for each integration variable, coming only from positively charged chirals under the $U(1)^{\mathrm{rk}\,G}$ Cartan of $G$.

\bibliographystyle{jhep}

\begin{thebibliography}{99}

\bibitem{Seiberg:1994pq}
N.~Seiberg, {\it {Electric - magnetic duality in supersymmetric nonAbelian
  gauge theories}},  {\em Nucl. Phys.} {\bf B435} (1995) 129--146,
  [\href{http://xxx.lanl.gov/abs/hep-th/9411149}{{\tt hep-th/9411149}}].

\bibitem{Hori:2006dk}
K.~Hori and D.~Tong, {\it {Aspects of Non-Abelian Gauge Dynamics in
  Two-Dimensional N=(2,2) Theories}},  {\em JHEP} {\bf 05} (2007) 079,
  [\href{http://xxx.lanl.gov/abs/hep-th/0609032}{{\tt hep-th/0609032}}].

\bibitem{Hori:2011pd}
K.~Hori, {\it {Duality In Two-Dimensional (2,2) Supersymmetric Non-Abelian
  Gauge Theories}},  {\em JHEP} {\bf 10} (2013) 121,
  [\href{http://xxx.lanl.gov/abs/1104.2853}{{\tt 1104.2853}}].

\bibitem{Pestun:2016zxk}
V.~Pestun {\em et~al.}, {\it {Localization techniques in quantum field
  theories}},  {\em J.\ Phys.\ A} {\bf 50} (2017), no.~44 440301,
  [\href{http://xxx.lanl.gov/abs/1608.02952}{{\tt 1608.02952}}].

\bibitem{Benini:2016qnm}
F.~Benini and B.~Le~Floch, {\it {Supersymmetric localization in two
  dimensions}},  {\em J. Phys. A} {\bf 50} (2017), no.~44 443003,
  [\href{http://xxx.lanl.gov/abs/1608.02955}{{\tt 1608.02955}}].

\bibitem{Aharony:2013dha}
O.~Aharony, S.~S. Razamat, N.~Seiberg, and B.~Willett, {\it {3d dualities from
  4d dualities}},  {\em JHEP} {\bf 07} (2013) 149,
  [\href{http://xxx.lanl.gov/abs/1305.3924}{{\tt 1305.3924}}].

\bibitem{Aharony:2013kma}
O.~Aharony, S.~S. Razamat, N.~Seiberg, and B.~Willett, {\it {3$d$ dualities
  from 4$d$ dualities for orthogonal groups}},  {\em JHEP} {\bf 08} (2013) 099,
  [\href{http://xxx.lanl.gov/abs/1307.0511}{{\tt 1307.0511}}].

\bibitem{Aganagic:2001uw}
M.~Aganagic, K.~Hori, A.~Karch, and D.~Tong, {\it {Mirror symmetry in
  (2+1)-dimensions and (1+1)-dimensions}},  {\em JHEP} {\bf 07} (2001) 022,
  [\href{http://xxx.lanl.gov/abs/hep-th/0105075}{{\tt hep-th/0105075}}].

\bibitem{Aharony:2017adm}
O.~Aharony, S.~S. Razamat, and B.~Willett, {\it {From 3d duality to 2d
  duality}},  {\em JHEP} {\bf 11} (2017) 090,
  [\href{http://xxx.lanl.gov/abs/1710.00926}{{\tt 1710.00926}}].

\bibitem{Aharony:2016jki}
O.~Aharony, S.~S. Razamat, N.~Seiberg, and B.~Willett, {\it {The long flow to
  freedom}},  {\em JHEP} {\bf 02} (2017) 056,
  [\href{http://xxx.lanl.gov/abs/1611.02763}{{\tt 1611.02763}}].

\bibitem{Gadde:2015wta}
A.~Gadde, S.~S. Razamat, and B.~Willett, {\it {On the reduction of 4d $
  \mathcal{N}=1 $ theories on $ {\mathbb{S}}^2 $}},  {\em JHEP} {\bf 11} (2015)
  163, [\href{http://xxx.lanl.gov/abs/1506.08795}{{\tt 1506.08795}}].

\bibitem{Gadde:2013lxa}
A.~Gadde, S.~Gukov, and P.~Putrov, {\it {(0, 2) trialities}},  {\em JHEP} {\bf
  03} (2014) 076, [\href{http://xxx.lanl.gov/abs/1310.0818}{{\tt 1310.0818}}].

\bibitem{Gadde:2014ppa}
A.~Gadde, S.~Gukov, and P.~Putrov, {\it {Exact Solutions of 2d Supersymmetric
  Gauge Theories}},  {\em JHEP} {\bf 11} (2019) 174,
  [\href{http://xxx.lanl.gov/abs/1404.5314}{{\tt 1404.5314}}].

\bibitem{Putrov:2015jpa}
P.~Putrov, J.~Song, and W.~Yan, {\it {(0,4) dualities}},  {\em JHEP} {\bf 03}
  (2016) 185, [\href{http://xxx.lanl.gov/abs/1505.07110}{{\tt 1505.07110}}].

\bibitem{Gukov:2019lzi}
S.~Gukov, D.~Pei, and P.~Putrov, {\it {Trialities of minimally supersymmetric
  2d gauge theories}},  {\em JHEP} {\bf 04} (2020) 079,
  [\href{http://xxx.lanl.gov/abs/1910.13455}{{\tt 1910.13455}}].

\bibitem{Csaki:1996eu}
C.~Csaki, W.~Skiba, and M.~Schmaltz, {\it {Exact results and duality for SP(2N)
  SUSY gauge theories with an antisymmetric tensor}},  {\em Nucl. Phys.} {\bf
  B487} (1997) 128--140, [\href{http://xxx.lanl.gov/abs/hep-th/9607210}{{\tt
  hep-th/9607210}}].

\bibitem{Hwang:2020wpd}
C.~Hwang, S.~Pasquetti, and M.~Sacchi, {\it {4d mirror-like dualities}},
  \href{http://xxx.lanl.gov/abs/2002.12897}{{\tt 2002.12897}}.

\bibitem{Dedushenko:2017osi}
M.~Dedushenko and S.~Gukov, {\it {IR duality in 2D $N=(0,2)$ gauge theory with
  noncompact dynamics}},  {\em Phys. Rev.} {\bf D99} (2019), no.~6 066005,
  [\href{http://xxx.lanl.gov/abs/1712.07659}{{\tt 1712.07659}}].

\bibitem{Tachikawa:2018sae}
Y.~Tachikawa, {\it {Lectures on 4d $N$=1 dynamics and related topics}},  2018.
\newblock \href{http://xxx.lanl.gov/abs/1812.08946}{{\tt 1812.08946}}.

\bibitem{Benini:2012cz}
F.~Benini and N.~Bobev, {\it {Exact two-dimensional superconformal R-symmetry
  and c-extremization}},  {\em Phys. Rev. Lett.} {\bf 110} (2013), no.~6
  061601, [\href{http://xxx.lanl.gov/abs/1211.4030}{{\tt 1211.4030}}].
  
  \bibitem{Benini:2013cda}
F.~Benini and N.~Bobev, {\it {Two-dimensional SCFTs from wrapped branes and
  c-extremization}},  {\em JHEP} {\bf 06} (2013) 005,
  [\href{http://xxx.lanl.gov/abs/1302.4451}{{\tt 1302.4451}}].

\bibitem{Gadde:2013wq}
A.~Gadde, S.~Gukov, and P.~Putrov, {\it {Walls, Lines, and Spectral Dualities
  in 3d Gauge Theories}},  {\em JHEP} {\bf 05} (2014) 047,
  [\href{http://xxx.lanl.gov/abs/1302.0015}{{\tt 1302.0015}}].

\bibitem{Gadde:2013ftv}
A.~Gadde and S.~Gukov, {\it {2d Index and Surface operators}},  {\em JHEP} {\bf
  03} (2014) 080, [\href{http://xxx.lanl.gov/abs/1305.0266}{{\tt 1305.0266}}].

\bibitem{Benini:2013nda}
F.~Benini, R.~Eager, K.~Hori, and Y.~Tachikawa, {\it {Elliptic genera of
  two-dimensional N=2 gauge theories with rank-one gauge groups}},  {\em Lett.
  Math. Phys.} {\bf 104} (2014) 465--493,
  [\href{http://xxx.lanl.gov/abs/1305.0533}{{\tt 1305.0533}}].

\bibitem{Benini:2013xpa}
F.~Benini, R.~Eager, K.~Hori, and Y.~Tachikawa, {\it {Elliptic Genera of 2d
  ${\mathcal{N}}$ = 2 Gauge Theories}},  {\em Commun. Math. Phys.} {\bf 333}
  (2015), no.~3 1241--1286, [\href{http://xxx.lanl.gov/abs/1308.4896}{{\tt
  1308.4896}}].

\bibitem{Closset:2013sxa}
C.~Closset and I.~Shamir, {\it {The $\mathcal{N}=1$ Chiral Multiplet on
  $T^2\times S^2$ and Supersymmetric Localization}},  {\em JHEP} {\bf 03}
  (2014) 040, [\href{http://xxx.lanl.gov/abs/1311.2430}{{\tt 1311.2430}}].

\bibitem{Benini:2015noa}
F.~Benini and A.~Zaffaroni, {\it {A topologically twisted index for
  three-dimensional supersymmetric theories}},  {\em JHEP} {\bf 07} (2015) 127,
  [\href{http://xxx.lanl.gov/abs/1504.03698}{{\tt 1504.03698}}].

\bibitem{Honda:2015yha}
M.~Honda and Y.~Yoshida, {\it {Supersymmetric index on $T^2 x S^2$ and elliptic
  genus}},  \href{http://xxx.lanl.gov/abs/1504.04355}{{\tt 1504.04355}}.

\bibitem{Witten:1988ze}
E.~Witten, {\it {Topological Quantum Field Theory}},  {\em Commun. Math. Phys.}
  {\bf 117} (1988) 353.

\bibitem{Witten:1991zz}
E.~Witten, {\em {Mirror manifolds and topological field theory}}, vol.~9,
  pp.~121--160.
\newblock 1998.
\newblock \href{http://xxx.lanl.gov/abs/hep-th/9112056}{{\tt hep-th/9112056}}.

\bibitem{Kutasov:2013ffl}
D.~Kutasov and J.~Lin, {\it {(0,2) Dynamics From Four Dimensions}},  {\em Phys.
  Rev.} {\bf D89} (2014), no.~8 085025,
  [\href{http://xxx.lanl.gov/abs/1310.6032}{{\tt 1310.6032}}].

\bibitem{tachi}
Y.~Tachikawa {\em unpublished}.

\bibitem{Intriligator:1995ne}
K.~A. Intriligator and P.~Pouliot, {\it {Exact superpotentials, quantum vacua
  and duality in supersymmetric SP(N(c)) gauge theories}},  {\em Phys. Lett.}
  {\bf B353} (1995) 471--476,
  [\href{http://xxx.lanl.gov/abs/hep-th/9505006}{{\tt hep-th/9505006}}].

\bibitem{10.1155/S1073792801000526}
J.~F. van Diejen and V.~P. Spiridonov, {\it {Elliptic Selberg integrals}},
  {\em International Mathematics Research Notices} {\bf 2001} (01, 2001)
  1083--1110.

\bibitem{2003math......9252R}
E.~M. {Rains}, {\it {Transformations of elliptic hypergometric integrals}},
  {\em arXiv Mathematics e-prints} (Sep, 2003) math/0309252,
  [\href{http://xxx.lanl.gov/abs/math/0309252}{{\tt math/0309252}}].

\bibitem{Benvenuti:2017lle}
S.~Benvenuti and S.~Giacomelli, {\it {Supersymmetric gauge theories with
  decoupled operators and chiral ring stability}},  {\em Phys. Rev. Lett.} {\bf
  119} (2017), no.~25 251601, [\href{http://xxx.lanl.gov/abs/1706.02225}{{\tt
  1706.02225}}].

\bibitem{Benvenuti:2018bav}
S.~Benvenuti, {\it {A tale of exceptional $3d$ dualities}},  {\em JHEP} {\bf
  03} (2019) 125, [\href{http://xxx.lanl.gov/abs/1809.03925}{{\tt
  1809.03925}}].

\bibitem{Amariti:2018wht}
A.~Amariti and L.~Cassia, {\it {USp(2N$_{c}$) SQCD$_{3}$ with antisymmetric:
  dualities and symmetry enhancements}},  {\em JHEP} {\bf 02} (2019) 013,
  [\href{http://xxx.lanl.gov/abs/1809.03796}{{\tt 1809.03796}}].

\bibitem{Pasquetti:2019uop}
S.~Pasquetti and M.~Sacchi, {\it {From 3$d$ dualities to 2$d$ free field
  correlators and back}},  {\em JHEP} {\bf 11} (2019) 081,
  [\href{http://xxx.lanl.gov/abs/1903.10817}{{\tt 1903.10817}}].

\bibitem{Pasquetti:2019tix}
S.~Pasquetti and M.~Sacchi, {\it {3d dualities from 2d free field correlators:
  recombination and rank stabilization}},  {\em JHEP} {\bf 01} (2020) 061,
  [\href{http://xxx.lanl.gov/abs/1905.05807}{{\tt 1905.05807}}].

\bibitem{4drankstab}
S.~Pasquetti and M.~Sacchi, {\it {4d rank stabilization duality}}, work
  in progress.

\bibitem{Aharony:1997gp}
O.~Aharony, {\it {IR duality in d = 3 N=2 supersymmetric USp(2N(c)) and U(N(c))
  gauge theories}},  {\em Phys. Lett.} {\bf B404} (1997) 71--76,
  [\href{http://xxx.lanl.gov/abs/hep-th/9703215}{{\tt hep-th/9703215}}].

\bibitem{Benini:2017dud}
F.~Benini, S.~Benvenuti, and S.~Pasquetti, {\it {SUSY monopole potentials in
  2+1 dimensions}},  {\em JHEP} {\bf 08} (2017) 086,
  [\href{http://xxx.lanl.gov/abs/1703.08460}{{\tt 1703.08460}}].
  
  \bibitem{Csaki:1997cu}
C.~Csaki, M.~Schmaltz, W.~Skiba, and J.~Terning, {\it {Selfdual N=1 SUSY gauge
  theories}},  {\em Phys. Rev. D} {\bf 56} (1997) 1228--1238,
  [\href{http://xxx.lanl.gov/abs/hep-th/9701191}{{\tt hep-th/9701191}}].

\bibitem{Pasquetti:2019hxf}
S.~Pasquetti, S.~S. Razamat, M.~Sacchi, and G.~Zafrir, {\it {Rank $Q$ E-string
  on a torus with flux}},  {\em SciPost Phys.} {\bf 8} (2020), no.~1 014,
  [\href{http://xxx.lanl.gov/abs/1908.03278}{{\tt 1908.03278}}].
  
  \bibitem{Garozzo:2020pmz}
I.~Garozzo, N.~Mekareeya, M.~Sacchi, and G.~Zafrir, {\it {Symmetry enhancement
  and duality walls in 5d gauge theories}},  {\em JHEP} {\bf 06} (2020) 159,
  [\href{http://xxx.lanl.gov/abs/2003.07373}{{\tt 2003.07373}}].
  
    \bibitem{HPS}
C.~Hwang, S.~Pasquetti, and M.~Sacchi, {\it {Flips, dualities and symmetry
  enhancements}},  \href{http://xxx.lanl.gov/abs/2010.10446}{{\tt 2010.10446}}.

\bibitem{Intriligator:1996ex}
K.~A. Intriligator and N.~Seiberg, {\it {Mirror symmetry in three-dimensional
  gauge theories}},  {\em Phys. Lett.} {\bf B387} (1996) 513--519,
  [\href{http://xxx.lanl.gov/abs/hep-th/9607207}{{\tt hep-th/9607207}}].

\bibitem{Kapustin:1999ha}
A.~Kapustin and M.~J. Strassler, {\it {On mirror symmetry in three-dimensional
  Abelian gauge theories}},  {\em JHEP} {\bf 04} (1999) 021,
  [\href{http://xxx.lanl.gov/abs/hep-th/9902033}{{\tt hep-th/9902033}}].

\bibitem{Kapustin:2010xq}
A.~Kapustin, B.~Willett, and I.~Yaakov, {\it {Nonperturbative Tests of
  Three-Dimensional Dualities}},  {\em JHEP} {\bf 10} (2010) 013,
  [\href{http://xxx.lanl.gov/abs/1003.5694}{{\tt 1003.5694}}].

\bibitem{Benvenuti:2016wet}
S.~Benvenuti and S.~Pasquetti, {\it {3d $ \mathcal{N} $ = 2 mirror symmetry,
  pq-webs and monopole superpotentials}},  {\em JHEP} {\bf 08} (2016) 136,
  [\href{http://xxx.lanl.gov/abs/1605.02675}{{\tt 1605.02675}}].
  
  \bibitem{Intriligator:2003jj}
K.~A. Intriligator and B.~Wecht, ``{The Exact superconformal R symmetry
  maximizes a},'' \href{http://dx.doi.org/10.1016/S0550-3213(03)00459-0}{{\em
  Nucl. Phys.} {\bfseries B667} (2003) 183--200},
\href{http://arxiv.org/abs/hep-th/0304128}{{\ttfamily arXiv:hep-th/0304128
  [hep-th]}}.

\bibitem{Spiridonov:2008zr}
V.~P. Spiridonov and G.~S. Vartanov, {\it {Superconformal indices for N = 1
  theories with multiple duals}},  {\em Nucl. Phys.} {\bf B824} (2010)
  192--216, [\href{http://xxx.lanl.gov/abs/0811.1909}{{\tt 0811.1909}}].

\bibitem{Dimofte:2012pd}
T.~Dimofte and D.~Gaiotto, {\it {An E7 Surprise}},  {\em JHEP} {\bf 10} (2012)
  129, [\href{http://xxx.lanl.gov/abs/1209.1404}{{\tt 1209.1404}}].

\bibitem{Razamat:2017hda}
S.~S. Razamat and G.~Zafrir, {\it {E$_{8}$ orbits of IR dualities}},  {\em
  JHEP} {\bf 11} (2017) 115, [\href{http://xxx.lanl.gov/abs/1709.06106}{{\tt
  1709.06106}}].

\bibitem{Razamat:2017wsk}
S.~S. Razamat, O.~Sela, and G.~Zafrir, {\it {Between Symmetry and Duality in
  Supersymmetric Quantum Field Theories}},  {\em Phys. Rev. Lett.} {\bf 120}
  (2018), no.~7 071604, [\href{http://xxx.lanl.gov/abs/1711.02789}{{\tt
  1711.02789}}].

\bibitem{Benini:2018bhk}
F.~Benini and S.~Benvenuti, {\it {$N=1$ QED in 2+1 dimensions: Dualities and
  enhanced symmetries}},  \href{http://xxx.lanl.gov/abs/1804.05707}{{\tt
  1804.05707}}.

\bibitem{Razamat:2018gbu}
S.~S. Razamat, O.~Sela, and G.~Zafrir, {\it {Curious patterns of IR symmetry
  enhancement}},  {\em JHEP} {\bf 10} (2018) 163,
  [\href{http://xxx.lanl.gov/abs/1809.00541}{{\tt 1809.00541}}].

\bibitem{PRS}
S.~S. Razamat, S.~Pasquetti, and M.~Sacchi {\em work in progress}.

\bibitem{Dimofte:2017tpi}
T.~Dimofte, D.~Gaiotto, and N.~M. Paquette, {\it {Dual boundary conditions in
  3d SCFT's}},  {\em JHEP} {\bf 05} (2018) 060,
  [\href{http://xxx.lanl.gov/abs/1712.07654}{{\tt 1712.07654}}].

\bibitem{Schellekens:1986yj}
A.~Schellekens and N.~Warner, {\it {Anomaly Cancellation and Selfdual
  Lattices}},  {\em Phys. Lett. B} {\bf 181} (1986) 339--343.

\bibitem{Schellekens:1986yi}
A.~Schellekens and N.~Warner, {\it {Anomalies and Modular Invariance in String
  Theory}},  {\em Phys. Lett. B} {\bf 177} (1986) 317--323.

\bibitem{Pilch:1986en}
K.~Pilch, A.~Schellekens, and N.~Warner, {\it {Path Integral Calculation of
  String Anomalies}},  {\em Nucl. Phys. B} {\bf 287} (1987) 362--380.

\bibitem{Witten:1986bf}
E.~Witten, {\it {Elliptic Genera and Quantum Field Theory}},  {\em Commun.
  Math. Phys.} {\bf 109} (1987) 525.

\bibitem{Witten:1987cg}
E.~Witten, {\it {The index of the Dirac operator in loop space}},  {\em Lect.
  Notes Math.} {\bf 1326} (1988) 161--181.

\bibitem{1993alg.geom..7001J}
L.~C. {Jeffrey} and F.~C. {Kirwan}, {\it {Localization for nonabelian group
  actions}},  {\em arXiv e-prints} (July, 1993) alg--geom/9307001,
  [\href{http://xxx.lanl.gov/abs/alg-geom/9307001}{{\tt alg-geom/9307001}}].

\end{thebibliography}

\end{document}